\documentclass[reqno]{amsart}
\usepackage[pagewise]{lineno}\linenumbers
\usepackage{graphicx}
\usepackage{caption}
\usepackage[labelformat=simple,labelfont={}]{subcaption}
\usepackage{bm}
\usepackage{amsmath}
\usepackage{mathtools}
\usepackage{eqparbox}
\usepackage{color}
\usepackage{multirow}
\usepackage{placeins}
\usepackage{amsmath}
\usepackage{nameref,hyperref}
\usepackage{xurl} 
\usepackage{epstopdf}

\newtheorem{theorem}{Theorem}[section]

\theoremstyle{definition}

\newtheorem{example}[theorem]{Example}
\theoremstyle{remark}

\numberwithin{equation}{section}

\newcommand{\dru}{\delta_{ru}}
\newcommand{\dur}{\delta_{ur}}
\newcommand{\mm}{\mu_u}
\newcommand{\mr}{\mu_r}
\newcommand{\bu}{\beta_u}
\newcommand{\br}{\beta_r}
\newcommand{\gu}{\gamma_u}
\newcommand{\gr}{\gamma_r}
\newcommand{\pu}{\rho_u}
\newcommand{\pr}{\rho_r}

\newcommand{\bx}{\bm{X}}

\title{Stochastic two-patch epidemic model with nonlinear recidivism}

\author[J.G. Calvo]{Juan G. Calvo}
\address{Centro de Investigación en Matemática Pura y Aplicada - Escuela de Matemática, Universidad de Costa Rica, San Jose, Costa Rica}
\email{juan.calvo@ucr.ac.cr}

\author[M.I. Simoy]{Mario I. Simoy}
\address{Instituto PLADEMA, Universidad Nacional del Centro de la Provincia de Buenos Aires, Paraje Arroyo Seco s/n, 7000 Tandil, Argentina}
\address{Instituto Multidisciplinario sobre Ecosistemas y Desarrollo Sustentable, Universidad Nacional del Centro de la Provincia de Buenos Aires, Paraje Arroyo Seco s/n, 7000 Tandil, Argentina}
\email{ignacio.simoy@gmail.com}

\author[J.P. Aparicio]{Juan P. Aparicio}
\address{Instituto de Investigaciones en Energ\'ia no Convencional (INENCO),  Consejo Nacional de Investigaciones Cient\'ificas y T\'ecnicas (CONICET), Universidad Nacional de Salta, Av. Bolivia 5150, 4400 Salta, Argentina.}
\email{juan.p.aparicio@gmail.com}

\author[J.E. Chac\'on]{José Emmanuel Chacón}
\address{Banco Nacional de Costa Rica, San Jose, Costa Rica}
\email{jechaconch@bncr.fi.cr}

\author[F. Sanchez]{Fabio Sanchez}
\address{Centro de Investigación en Matemática Pura y Aplicada - Escuela de Matemática, Universidad de Costa Rica, San Jose, Costa Rica}
\email{fabio.sanchez@ucr.ac.cr}

\begin{document}
\nolinenumbers
\subjclass[2020]{Primary 60J20, 92D30; Secondary 92-08, 60H15} 


\keywords{Population dynamics, mathematical modeling, epidemic model, population movement, stochastic model}

\begin{abstract}
We develop a stochastic two-patch epidemic model with nonlinear recidivism to investigate infectious disease dynamics in heterogeneous populations. Extending a deterministic framework, we introduce stochasticity to account for random transmission, recovery, and inter-patch movement fluctuations. We showcase the interplay between local dynamics and migration effects on disease persistence using Monte Carlo simulations and three stochastic approximations—discrete-time Markov chain (DTMC), Poisson, and stochastic differential equations (SDE). Our analysis shows that stochastic effects can cause extinction events and oscillations near critical thresholds like the {\it basic reproduction number}, \(\mathcal{R}_0\), phenomena absent in deterministic models. Numerical simulations highlight source-sink dynamics, where one patch is a persistent infection source while the other experiences intermittent outbreaks.

\end{abstract}

\maketitle

\allowdisplaybreaks 

\section{Introduction}

Epidemic models have been used extensively to study infectious diseases \cite{brauer2001,chowell2006}. However, many unpredictable factors can change the course of an epidemic. The COVID-19 pandemic is a recent one that affected every country in the world~\cite{who2020,cucinotta2020}. Many, if not all, of these countries had to implement containment measures to slow down the spread of what was at the time the novel SARS-CoV-2 virus~\cite{anderson2020,maxmen2020}. Some measures, like lockdowns, were extreme measures limiting population movement~\cite{who2020,cdc2020,cohen2020}.

The study of infectious diseases and their spread has long been a topic of interest for researchers and policymakers alike. The complexity of these diseases and their transmission dynamics can make them difficult to understand and predict, making it essential to develop mathematical models that can help improve our understanding and ability to control outbreaks. Stochastic epidemic models are a mathematical framework used to study infectious diseases and can provide valuable insights into the spread of diseases and the impact of control strategies.

Movement between populations has been at the forefront of epidemic modeling in recent years, given the relative ease with which individuals can move between towns, cities, countries, and continents. The COVID-19 pandemic highlighted the importance of social determinants in the spread of infectious diseases~\cite{delvalle2005}, particularly population movement~\cite{garcia2022}. Different mathematical models used during the pandemic incorporated these effects and showed promising results~\cite{sanchez2022,dazatorres2022,valles2022}.

Moreover, {\it temporal movement} can be defined as individuals traveling to a distinct location from their origin and returning to their original location within a usually short temporal period. For example, individuals living in a suburban area move to the central city to study or work but return to their homes at the end of the day. On the other hand, {\it migration} is a long-term movement. Individuals who move from one residence to another remain in the new site for long periods. 



Here, we study the disease dynamics in two patches using a stochastic approach with population movement. Taking as a starting point the deterministic model developed in~\cite{calvo2019}, we incorporate the stochasticity of each event with its corresponding probability of occurrence. In addition to the stochastic Montecarlo model, we develop three different approximations for the stochastic model (discrete-time Markov chain model, Poisson approximation, and stochastic differential equations) to compare the dynamic behavior between them and with the deterministic model. A brief review of the main results of the deterministic case~\cite{calvo2019} is presented in the next section.

The article is organized as follows. In Section \ref{sec:model}, we present the two-patch stochastic model. Three approximations to the stochastic model are presented in Section \ref{sec:approx}. Numerical experiments are then included in Section \ref{sec:numericalExp}, and some final remarks and conclusions are discussed in Section \ref{sec:conclusions}.

\section{Two-patch stochastic model} \label{sec:model}
We consider a two-patch stochastic model incorporating nonlinear recidivism based on the deterministic model studied in~\cite{calvo2019}. We consider two patches, urban and rural areas, denoted by the subscripts $u$ and $r$, respectively. For each patch, we develop a compartmental model with three classes: susceptible, infected, and temporarily recovered (but still susceptible to re-infection), denoted by $S_j$, $I_j$ and $\widetilde{S}_j$, $j\in\lbrace u,r\rbrace$, respectively. Movement between patches is modeled via the functions $\delta_{ij}(t)$, which denote the fraction of individuals who travel from patch $i \in\lbrace u,r\rbrace$ to patch $j \in\lbrace u,r\rbrace$ (with $i \neq j$) at time $t$. We consider the model given by the following nonlinear system of differential equations:

\begin{equation} \label{eq:sys}
\begin{array}{rcl}
\dfrac{d S_u}{dt} &=& \Lambda_u - \beta_u \dfrac{I_u}{N_u} S_u -\mu_u S_u + \dru S_r - \dur S_u, \vspace{.2cm}\\
\dfrac{d I_u}{dt} &=& \beta_u \dfrac{I_u}{N_u} S_u - (\mu_u+\gamma_u)I_u + \rho_u \dfrac{I_u}{N_u} \widetilde{S}_u +\dru I_r - \dur I_u,\vspace{.2cm}\\
\dfrac{d \widetilde{S}_u}{dt} &=& \gamma_u I_u - \rho_u \dfrac{I_u}{N_u} \widetilde{S}_u  - \mu_u \widetilde{S}_u +\dru \widetilde{S}_r - \dur \widetilde{S}_u,\vspace{.4cm}\\
\dfrac{d S_r}{dt} &=& \Lambda_r - \beta_r \dfrac{I_r}{N_r} S_r -\mu_r S_r + \dur S_u - \dru S_r,\vspace{.2cm}\\
\dfrac{d I_r}{dt} &=& \beta_r \dfrac{I_r}{N_r} S_r - (\mu_r+\gamma_r)I_r + \rho_r \dfrac{I_r}{N_r} \widetilde{S}_r +\dur I_u - \dru I_r,\vspace{.2cm}\\
\dfrac{d \widetilde{S}_r}{dt} &=& \gamma_r I_r - \rho_r \dfrac{I_r}{N_r} \widetilde{S}_r  - \mu_r \widetilde{S}_r +\dur \widetilde{S}_u - \dru \widetilde{S}_r,
\end{array}
\end{equation}
with appropriate initial conditions, where $N_u := S_u+I_u+\widetilde{S}_u$ and $N_r := S_r+I_r+\widetilde{S}_r$ denote the total population of the urban and rural patch, respectively, {
and $\Lambda_j$, $\mu_j$, $\beta_j$, $\gamma_j$ and $\rho_j$ denote the recruitment, mortality, infection, recovery, and relapse rates, respectively}. The dynamics of the total population is given by 

\begin{equation} \label{eq:sysN}
\begin{array}{rcl}
\dfrac{d N_u}{dt} &=& \Lambda_u -\mu_u N_u -\dur N_u+\dru N_r,\vspace{.2cm}\\
\dfrac{d N_r}{dt} &=& \Lambda_r -\mu_r N_r -\dru N_r+\dur N_u,
\end{array}
\end{equation}
and the equilibrium point of system \eqref{eq:sysN} is given by 

\begin{equation*}
\begin{array}{rcl}
N_u^*=\dfrac{\mu_r \Lambda_u+(\Lambda_u+\Lambda_r)\delta_{ru}}{(\mu_r+\delta_{ru})\mu_u+\delta_{ur}\mu_r}, \vspace{.2cm}\\
N_r^*= \dfrac{ \mu_u \Lambda_r +  (\Lambda_u+\Lambda_r)\delta_{ur}}{(\mu_r+\delta_{ru})\mu_u+\delta_{ur}\mu_r}.
\end{array}
\end{equation*}

Deterministic disease dynamics in the case of isolated populations ($\dur=\dru=0$) are described by the well-known system

\begin{equation} \label{eq:sysOnePatch}
\begin{array}{rcl}
\dfrac{d S}{dt} &=& \Lambda - \beta \dfrac{I}{N} S -\mu S, \\ \\
\dfrac{d I}{dt} &=& \beta \dfrac{I}{N} S- (\mu+\gamma)I + \rho\dfrac{I}{N} \widetilde{S},\\ \\
\dfrac{d \widetilde{S}}{dt} &=& \gamma I - \rho \dfrac{I}{N} \widetilde{S} - \mu\widetilde{S},
\end{array}
\end{equation}
with basic reproduction number $\mathcal{R}_{0}=\beta/(\gamma+\mu)$. 
This system has a backward bifurcation and can present a positive, stable, steady state despite $\mathcal{R}_{0}<1$, depending on the initial conditions; see \cite{sanchez2007}.  Meanwhile, $\mathcal{R}_{0}>1$ guarantees the existence of a positive, stable, steady state (commonly known as the endemic state or endemic equilibrium).

When the two populations are connected by migration as given in \eqref{eq:sys}, the disease-free state is asymptotically stable if

\begin{subequations}\label{eq:condR0}
\begin{align}
	\label{lem:symsta_a} \mathcal{R}_{0u} &< 1 + \dfrac{\dur}{\mm+\gu}\,,\\ 
    \label{lem:symsta_b} \mathcal{R}_{0r} &< 1 + \dfrac{\dru}{\mr+\gr},\\ 
    \label{lem:symsta_c} \dfrac{\dur}{\mm+\gu}\dfrac{\dru}{\mr+\gr} &< \left(\mathcal{R}_{0u}-1-\dfrac{\dur}{\mm+\gu}\right)\left(\mathcal{R}_{0r}-1-\dfrac{\dru}{\mr+\gr}\right)\,,
\end{align}
\end{subequations}
where 
\begin{equation*}
	\mathcal{R}_{0u} := \dfrac{\bu}{\gu+\mm}\, \quad \text{and} \quad \mathcal{R}_{0r} := \dfrac{\br}{\gr+\mr}\,
\end{equation*}	
are  the \textit{local basic reproduction number} (in absence of migration); see \cite[Lemma 3.1]{calvo2019}.

On the other hand, if
$\mathcal{R}_{0u} > 1+\dfrac{\dur}{\mm+\gu}$ and $\mathcal{R}_{0r} > 1+\dfrac{\dru}{\mr+\gr}$, there exists at least one endemic state for which $I_u^*>0$ and $I_r^*>0$ (\cite[Lemma 3.7]{calvo2019}); i.e., there are infected individuals at steady state in both urban and rural populations. We remark that System \eqref{eq:sys} can have at most eight positive and feasible fixed points; see \cite[Example 4.7]{calvo2019}.

The main hypotheses behind model \eqref{eq:sys} are the following: (i) populations are considered completely homogeneous, (ii) there is homogeneous mixing, (iii) the probability of {\it infection, recovery, and relapse} is constant (independent of the age of infection), and (iv) mortality {and recruitment are} constant (independent of the age of the individuals).

With the same assumptions, the corresponding stochastic model is defined for all the possible events and their corresponding probabilities of occurrence. For example, the probability of  recruitment of a susceptible individual in the compartment $S_j$ ($j\in\{u, r\}$) in an infinitesimal period $\delta t$ is

$$P(S_j\rightarrow S_j+1)= \Lambda_j \delta t + o(\delta t) \quad \text{where} \quad \lim \frac{o(\delta t)}{\delta t}=0.$$


Probabilities per unit of time are the transition rates ($\Lambda_j$ in the example above) and completely determine the stochastic process; see Table \ref{table:transitions}.

\begin{table}[h]\begin{center}
\begin{tabular}{c c c}
Event & Transition rate & Effect\\ \hline
Recruitment & $\Lambda_j$ & $S_j\rightarrow S_j+1$\\\hline
Death & $\mu_j X_j$ & $X_j \rightarrow X_j - 1$ \\\hline
Infection & $\beta_j \frac{I_j}{N_j} S_j$ & {\begin{tabular}[c]{@{}l@{}} $S_j\rightarrow S_j-1$\\ $I_j\rightarrow I_j+1$\end{tabular}} \\ \hline
Recidivism &$ \rho_j \frac{I_j}{N_j} \widetilde{S_j} $  & {\begin{tabular}[c]{@{}l@{}} $I_j\rightarrow I_j+1$\\ $\widetilde{S}_j\rightarrow \widetilde{S}_j-1$ \end{tabular}} \\ \hline
Recovery & $\gamma_jI_j$ & {\begin{tabular}[c]{@{}l@{}} $I_j\rightarrow I_j-1$\\ $\widetilde{S}_j\rightarrow \widetilde{S}_j+1$ \end{tabular}} \\ \hline
Movement & $\delta_{ij} X_i$ & {\begin{tabular}[c]{@{}l@{}}  $X_i \rightarrow X_i - 1$\\ $X_j \rightarrow X_j + 1$   \end{tabular}} \\ \hline
\end{tabular}
\end{center}
\caption{ Events, transition rates and their corresponding effects on populations for the stochastic process considering $i,j \in \{u,r\}$ with $i \neq j$ and $X \in \{S, I, \widetilde{S}\}$. \label{table:transitions}}
\end{table}

Montecarlo simulations are performed considering that the time interval between successive events is a random variable exponentially distributed with parameters equal to the sum of all the transition rates. The event is selected with probabilities proportional to the transition rates. 

\section{Approximations to the stochastic model} \label{sec:approx}

In this section, we include three variants for the stochastic model described in Section \ref{sec:model} to compare their behavior with the deterministic model.

\subsection{A discrete-time Markov chain model}

We first introduce a discrete-time Markov chain (DTMC) model for System \eqref{eq:sys}; see \cite[Chapter 2]{allen2003} for a complete study on these models. Let $S_j(t)$, $I_j(t)$, $\widetilde{S}_j(t)$ ($j\in \lbrace u,r\rbrace$) denote discrete random variables for the number of susceptible, infected, and temporarily recovered individuals at time $t$ in patch $j$. We assume that for $j\in \lbrace u,r\rbrace$ it holds that $$S_j(t), I_j(t), \widetilde{S}_j(t)\in \mathbb{N},\quad t\in\lbrace 0, t_1, t_2, \ldots \rbrace,$$
and that the model satisfies the Markov property; i.e., at time $t_j$ the process depends only on the state at previous time $t_{j-1}$. We also assume that the time step is sufficiently small such that at every time step, only one event occurs (recruitment, death, infection, re-infection, recovery, movement), with transition rate as given in Table \ref{table:transitions}.

\subsection{The Poisson approximation}

Poisson approximation is a fixed time step ($\delta t$) approximation that converges to the stochastic model (defined as in Table \ref{table:transitions}) when the time step goes to zero (see for example \cite{aparicio2001}). For the simpler case of one isolated patch (as in the deterministic version \eqref{eq:sysOnePatch}), $S$, $I$, and $\widetilde{S}$ are random variables where their values at time $t+\delta t$ are given by

\begin{equation*}
\begin{array}{rcl}
{S}(t+\delta t) &=&S(t) +n_{b}-n_{inf}-n_{mS}, \\
I(t+\delta t) &=&I(t) +n_{inf}+n_{recid}-n_{mI}-n_r, \\
\widetilde{S} (t+\delta t) &=&      \widetilde{S}(t)-n_{recid}-n_{m\widetilde{S}}+n_r, 
\end{array}
\end{equation*}
where $n_j$ are Poisson random variables with parameters equal to the corresponding transition rates (as defined in Table \ref{table:transitions}) multiplied by the time step. 
The random variables $n_b, n_{inf}, n_{mX}, n_{recid}, n_r$ denote the number of births, infections, deaths at compartment $X\in\{S,I,\widetilde{S}\}$, relapses and recoveries, respectively, during the time step $\delta t$. For example, $n_{inf}$ is a random variable with Poisson distribution and parameter $\beta\frac{I}{N} S \delta t$. 

\subsection{A stochastic differential equation model} \label{sec:sde}

Let 
$$\Delta \bx(t) = (\Delta S_u, \Delta I_u,\Delta \widetilde{S}_u, \Delta S_r, \Delta I_r,\Delta \widetilde{S}_r)^T$$ be the vector of change in the population at time $t$. The expected value  of $\Delta \bx(t)$ is

$$\mathbb{E}\big(\Delta \bx(t)\big) =
\begin{bmatrix}
     \Lambda_u- \frac{\beta_u\, I_u\, S_u}{N_u} - \mu_u\, S_u  + \delta_{ru}\, S_r-\delta_{ur}\, S_u  \\
    \dfrac{\beta_u\, I_u\, S_u}{N_u} - (\mu_u+\gamma_u)\, I_u - + \dfrac{\rho_u\, I_u\, \widetilde{S}_u}{N_u}  + \delta_{ru}\, I_r -\delta_{ur}\, I_u  \\
     \gamma_u\, I_u - \dfrac{\rho_u\, I_u\, \widetilde{S}_u}{N_u} - \mu_u\, \widetilde{S}_u + \delta _{ru}\, \widetilde{S}_r-\delta _{ur}\, \widetilde{S}_u  \\
    \Lambda_r  - \dfrac{\beta_r\, I_r\, S_r}{N_r} - \mu_r\, S_r  + \delta_{ur}\, S_u - \delta_{ru}\, S_r \\
    \dfrac{\beta_r\, I_r\, S_r}{N_r} - ( \mu_r+\gamma_r)\, I_r\, + \dfrac{\rho_r\, I_r\, \widetilde{S}_r}{N_r}  +\delta_{ur}\, I_u - \delta_{ru}\, I_r\\
    \gamma_r\, I_r  - \dfrac{\rho_r\, I_r\, \widetilde{S}_r}{N_r}  - \mu_r\, \widetilde{S}_r + \delta _{ur}\, \widetilde{S}_u - \delta _{ru}\,  \widetilde{S}_r
\end{bmatrix} \Delta t.$$
The covariance matrix of $\Delta \bx(t)$ is 
$$\mathbb{V}\big(\Delta \bx(t)\big) = \mathbb{E}\big(\Delta \bx(t) \;\Delta \bx(t)^{T}\big) - \mathbb{E}\big(\Delta \bx(t)\big)\;\mathbb{E}\big(\Delta \bx(t)\big)^{T} \approx \mathbb{E}\big(\Delta \bx(t) \;\Delta \bx(t)^{T}\big),$$
where we drop the second order terms that are $o([\Delta t]^2)$.  Note that $\Delta \bx(t) \;\Delta \bx(t)^{T}$ is a $6\times6$ matrix, and $\mathbb{V}\big(\Delta \bx(t)\big)$ is the matrix whose entries are the expected values of the entries of $\Delta \bx(t) \;\Delta \bx(t)^{T}$. An explicit formula for 
$\mathbb{V}\big(\Delta \bx(t)\big)$ is shown in Appendix \ref{ap:expectedValueCovariance}.

We assume that $\Delta \bx(t)$ has an approximate multivariate normal distribution for small $\Delta t$; i.e., $\Delta \bx(t) \sim \mathcal{N}\big(\mathbb{E}\big(\Delta \bx(t)\big),\,\mathbb{V}\big(\Delta \bx(t)\big)\big)$, and therefore, the random vector $\bx(t + \Delta t)$ can be approximated as
$$\bx(t + \Delta t) = \bx(t) + \Delta \bx(t) \approx \bx(t) + \mathbb{E}\big(\Delta \bx(t)\big) + \sqrt{\mathbb{V}\big(\Delta \bx(t)\big)} \bm{\eta},$$
where $\bm{\eta} \sim \mathcal{N}(\bm{0}_{6 \times 1},\,I_{6 \times 6})$. Note that as the covariance matrix $\mathbb{V}\big(\Delta \bx(t)\big)$ is symmetric and positive definite, it has a unique square root.



The Stochastic Differential Equation (SDE) system is then given by:

\begin{align}
    dS_u =& (\Lambda_u - \beta_u \dfrac{I_u}{N_u} S_u -\mu_u S_u + \dru S_r - \dur S_u)dt + \sqrt{\Lambda_u }\ dW_{10} -\sqrt{\mu_uS_u }\ dW_{11}\nonumber \\
    &+ \sqrt{\dru S_r}\ dW_2 - \sqrt{\beta_u \dfrac{I_u}{N_u} S_u}\ dW_7 -\sqrt{\dur S_u}\ dW_1 \nonumber,\\
    dI_u =& (\beta_u \dfrac{I_u}{N_u} S_u - (\mu_u+\gamma_u)I_u + \rho_u \dfrac{I_u}{N_u} \widetilde{S}_u +\dru I_r - \dur I_u)dt -\sqrt{\dur I_u}\ dW_3 \nonumber \\
    &+ \sqrt{\dru I_r}\ dW_4+\sqrt{\beta_u \dfrac{I_u}{N_u} S_u} \ dW_7 +\sqrt{\rho_u \dfrac{I_u}{N_u} \widetilde{S}_u}\ dW_8\nonumber \\
    &-\sqrt{\gamma_u I_u}\ dW_9 - \sqrt{\mm I_u}\ dW_{12}, \nonumber \\
    d\widetilde{S}_u =& (\gamma_u I_u - \rho_u \dfrac{I_u}{N_u} \widetilde{S}_u  - \mu_u \widetilde{S}_u +\dru \widetilde{S}_r - \dur \widetilde{S}_u)dt -\sqrt{\dur \widetilde{S}_u}\ dW_5 \nonumber \\
    & +\sqrt{\dru \widetilde{S}_r}\ dW_6 -\sqrt{\rho_u \dfrac{I_u}{N_u} \widetilde{S}_u} \ dW_8+\sqrt{\gamma_u I_u}\ dW_9-\sqrt{\mm \widetilde{S}_u}\ dW_{13} \nonumber, \\
    dS_r =& (\Lambda_r - \beta_r \dfrac{I_r}{N_r} S_r -\mu_r S_r + \dur S_u - \dru S_r)dt +\sqrt{\Lambda_r }\ dW_{17}  - \sqrt{\dru S_r}\ dW_2 \nonumber \\
    &+\sqrt{\dur S_u}\ dW_1 - \sqrt{\beta_r \dfrac{I_r}{N_r} S_r}\ dW_{14}-\sqrt{ \mu_r S_r}\ dW_{18}, \nonumber \\
    dI_r =& (\beta_r \dfrac{I_r}{N_r} S_r - (\mu_r+\gamma_r)I_r + \rho_r \dfrac{I_r}{N_r} \widetilde{S}_r +\dur I_u - \dru I_r)dt +\sqrt{\beta_r \dfrac{I_r}{N_r} S_r}\ dW_{14}\nonumber \\
    &- \sqrt{\mu_r I_r}\ dW_{19}-\sqrt{\gamma_rI_r}\ dW_{16} +\sqrt{ \rho_r \dfrac{I_r}{N_r} \widetilde{S}_r }\ dW_{15}\nonumber \\
    &+\sqrt{\dur I_u }\ dW_3 -\sqrt{ \dru I_r}\ dW_4, \nonumber\\
    d\widetilde{S}_r =& (\gamma_r I_r - \rho_r \dfrac{I_r}{N_r} \widetilde{S}_r  - \mu_r \widetilde{S}_r +\dur \widetilde{S}_u - \dru \widetilde{S}_r)dt + \sqrt{\gamma_r I_r } \hspace{0.1cm} dW_{16} \nonumber \\
    &- \sqrt{\rho_r \dfrac{I_r}{N_r} \widetilde{S}_r }\hspace{0.1cm} dW_{15}  - \sqrt{\mu_r \widetilde{S}_r }\hspace{0.1cm} dW_{20} +\sqrt{\dur \widetilde{S}_u}\hspace{0.1cm} dW_5 - \sqrt{\dru \widetilde{S}_r } \hspace{0.1cm} dW_6. \nonumber
\end{align}

\noindent
where $W_i (1\leq i\leq 20)$ are independent Wiener processes. We refer to Appendix \ref{ap:numerical_scheme} for describing the numerical scheme used to solve the SDE system.

\section{Numerical Experiments} \label{sec:numericalExp}

In this section, we explore four examples to compare the behavior of the deterministic model and the stochastic variants we have discussed.

\subsection{Comparing deterministic and stochastic dynamics}

We compare the numerical results for the deterministic model as presented in \cite{calvo2019}, the stochastic model presented in Section \ref{sec:model}, and the approximations considered in Section \ref{sec:approx}. In all cases, rates $\mu$, $\beta$, $\gamma$ and $\rho$ are expressed in day$^{-1}$.

\begin{example}\label{ex:1}
We first consider the effect of having movement only in one direction, from the rural patch to the urban one ($\dur=0, \dru > 0$); see \cite[Example 1]{calvo2019}. 
We consider the following parameter values:

\begin{equation*}
	\begin{array}{ccccc}
\mm = 1/(365\cdot 80)\,, & \pu = 0.08\,, & \gu = 0.01\,, & \bu = 0.03\,, & \Lambda_u = \mm N_{u0},\\
\mr = 1/(365\cdot 70)\,, & \pr = 0.04\,, & \gr = 0.01\,, & \br = 0.02\,, & \Lambda_r = \mr N_{r0},
\end{array}
\end{equation*}
with initial conditions 
\begin{equation*}
	\begin{array}{ccc}
S_{u0} = 999\,, & I_{u0} = 1\,, & \widetilde{S}_{u0} = 0\,,\\
S_{r0} = 300\,, & I_{r0} = 0\,, & \widetilde{S}_{r0} = 0\,.
\end{array}
\end{equation*}
In this case,
\begin{align*}
	\mathcal{R}_{0u} \approx 3.00 \quad \text{and} \quad \mathcal{R}_{0r} \approx 2.00 \,,
\end{align*}
and we have a positive endemic state for the deterministic model in the urban patch. The population in rural areas will decrease to zero. According to the approximations for the stochastic model, we simulated the infectious dynamics using DTMC, Poisson, and SDE models; see Figure~\ref{fig:ex1b} where we present ten different simulations.

\begin{figure}[ht!]
\centering
\includegraphics[width=0.49\linewidth]{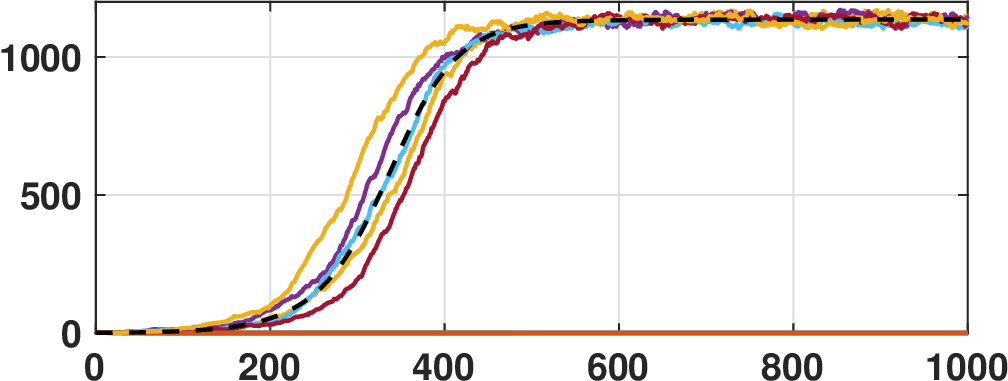} 
\hfill
\includegraphics[width=0.49\linewidth]{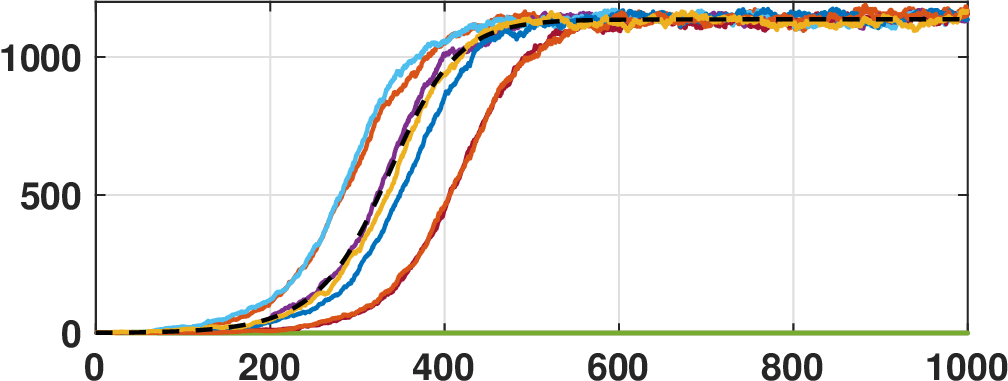} 
\hfill
\includegraphics[width=0.49\linewidth]{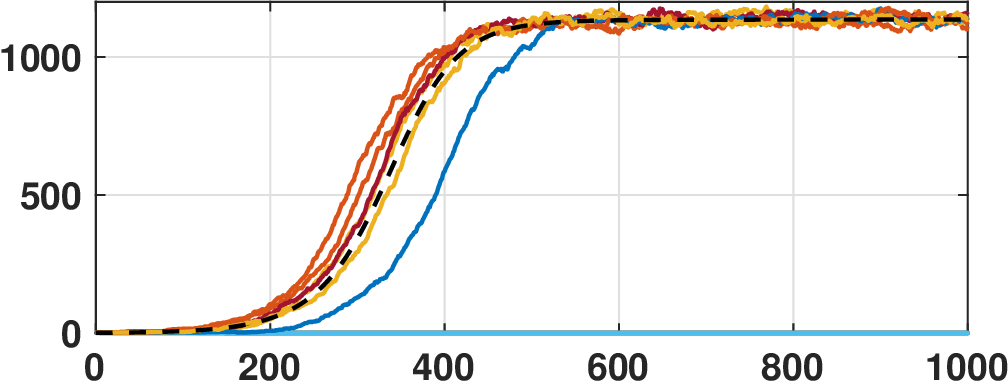} 
\hfill
\includegraphics[width=0.49\linewidth]{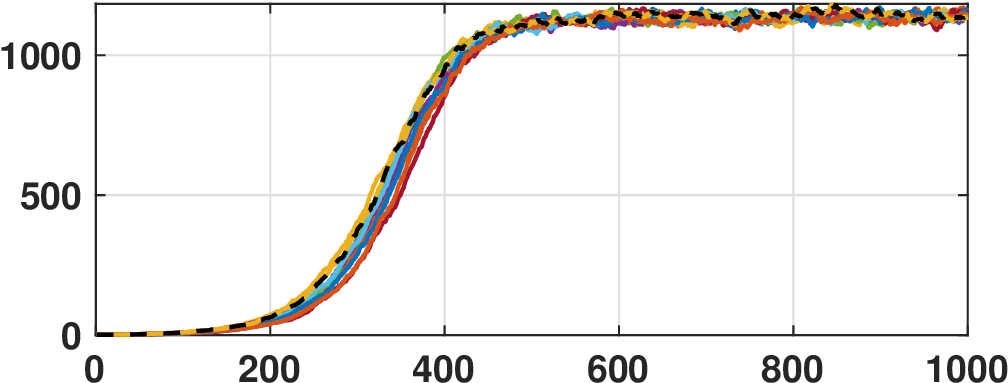} 
\caption{\label{fig:ex1b} $I_u$ as a function of time (days) for ten simulations with $\dur = 0$, $\dru = 0.01$ for the (upper left) stochastic, (upper right) DTMC, (lower left) Poisson approximation and (lower right) SDE models. For the 10 simulations, 5, 7, 6, and 10 curves reach a positive state (endemic equilibrium) for the stochastic, DTMC, Poisson, and SDE models, respectively. Hence, unlike the deterministic model, stochastic models can reach a disease-free state. 
The deterministic solution is shown in the black dashed line; see Example \ref{ex:1}.}
\end{figure}


Although the deterministic model guarantees a positive endemic state, our analysis of 1000 simulations revealed that $33.0\%$, $34.3\%$, and $31.8\%$ of the stochastic, DTMC, and Poisson approximation models, respectively, reached a disease-free state.
For the urban patch, Figure~\ref{fig:ex1b} shows that the SDE simulations always reach a positive endemic state. 
According to initial conditions and movement restrictions, no individuals in the rural patch were infected during the simulations.


In Figure \ref{fig:example1_1}, we show one stochastic curve $I_u(t)/N_u(t)$ for different values of $\dru$. The deterministic model reaches the same endemic state, with slight differences in the time it is reached. We observe no significant differences between the four models when they reach a positive state.

\begin{figure}[htb!]
\centering
\includegraphics[width=0.49\linewidth]{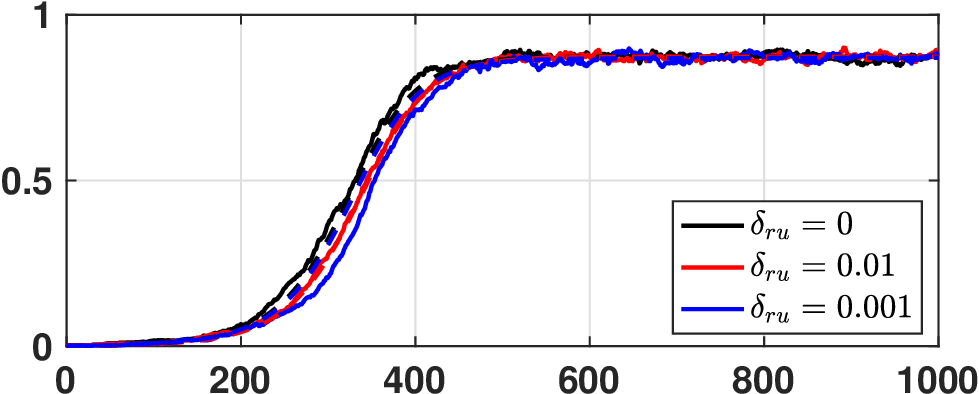}
\hfill
\includegraphics[width=0.49\linewidth]{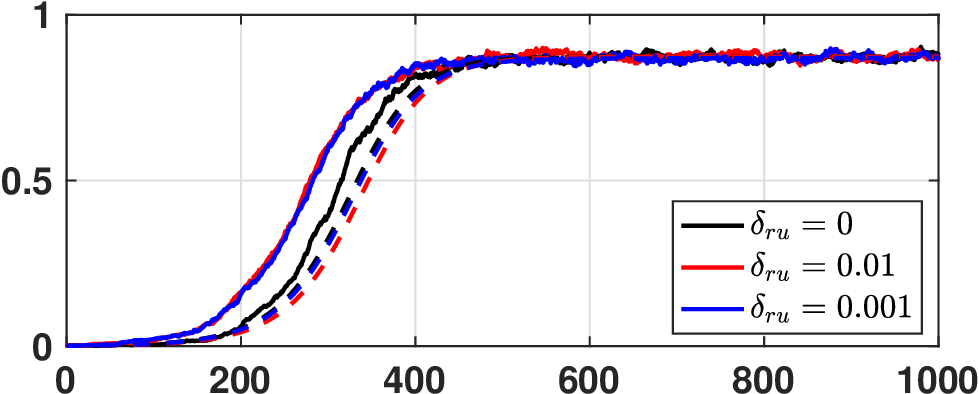}
\hfill
\includegraphics[width=0.49\linewidth]{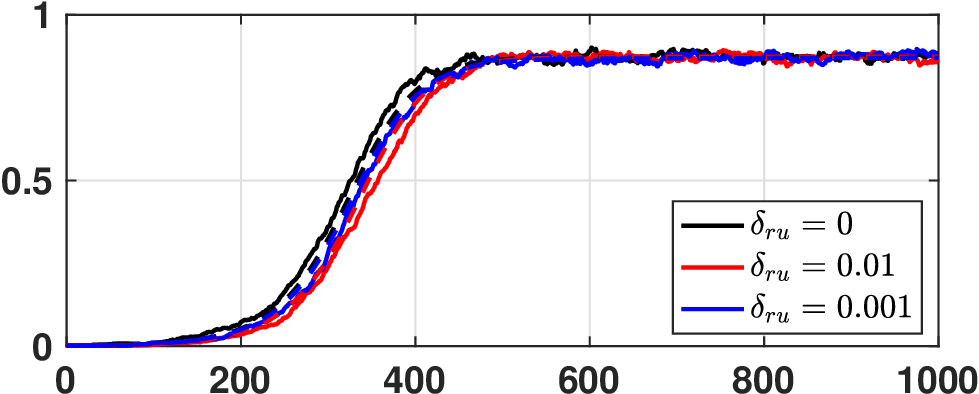}
\hfill
\includegraphics[width=0.49\linewidth]{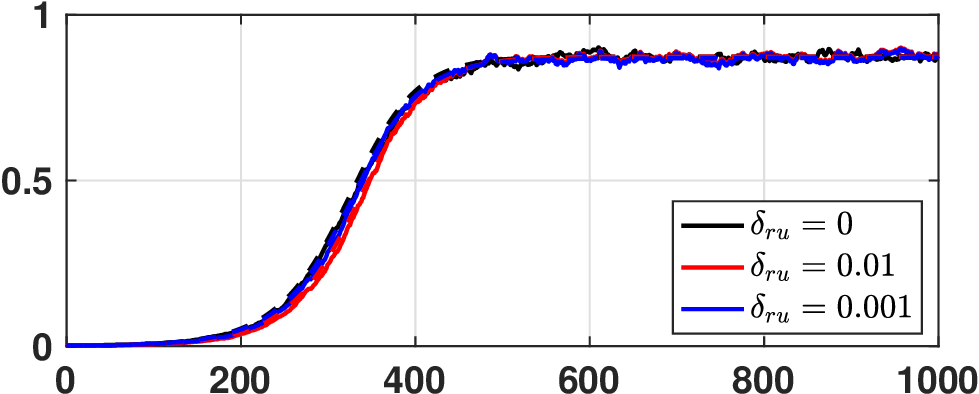}
\caption{\label{fig:example1_1} One stochastic curve $I_u(t)/N_u(t)$ as a function of time (days) for $\delta_{ur}=0$ and different values of $\delta_{ru}$, for (upper left) stochastic, (upper right) DTMC, (lower left) Poisson approximation and (lower right) SDE models; see Example \ref{ex:1}. Dashed lines correspond to the deterministic solution.}
\end{figure}

We finally show the solutions for the stochastic model 
for the urban and rural patches, $\dru = 0$ (i.e., when there is no movement) and $\dru = 0.01$, in Figure~\ref{fig:example1_2345}. We observe that population movement in one direction increases the peak in the number of infected individuals. The three approximations to the stochastic model follow the same dynamics observed in Figure~\ref{fig:example1_2345}.

\begin{figure}[htb!]
\centering 
\includegraphics[width=0.49\linewidth]{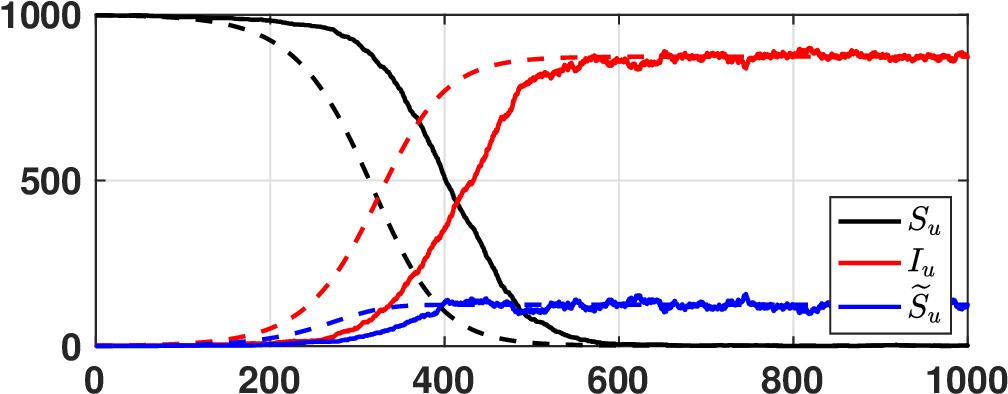}
\hfill
\includegraphics[width=0.49\linewidth]{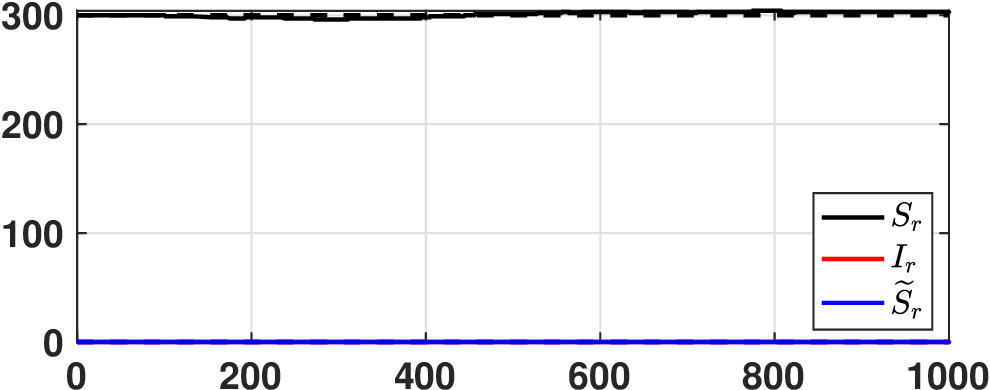}
\includegraphics[width=0.49\linewidth]{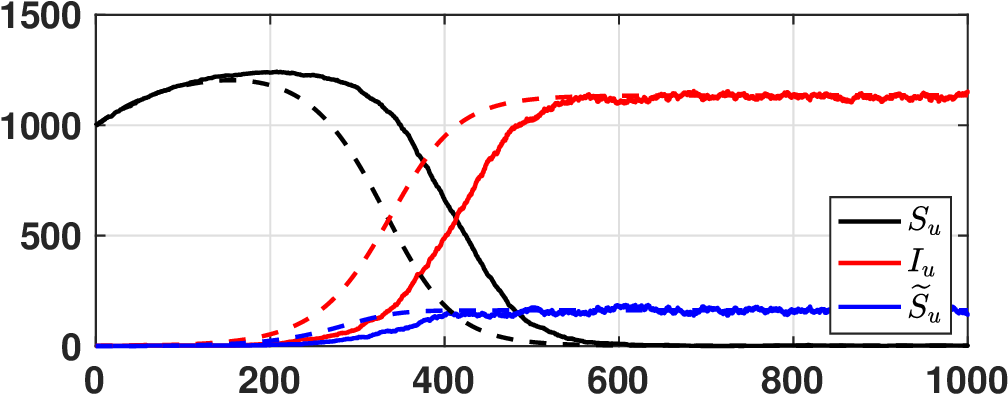}
\hfill
\includegraphics[width=0.49\linewidth]{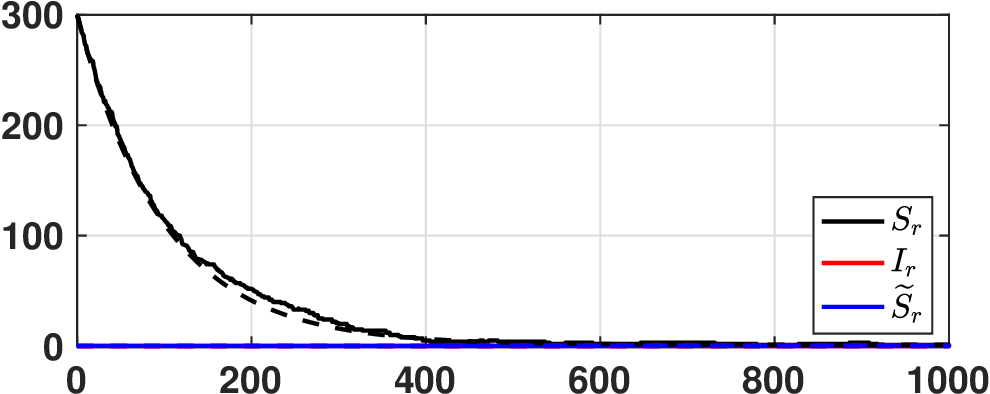}
\caption{\label{fig:example1_2345} Number of individuals for each compartment as a function of time (days). We show the effect of population movement in just one direction (from the rural patch to the urban one). In this scenario, $\delta_{ur} = 0$ and (top) $\delta_{ru} = 0$, (bottom) $\delta_{ru} = 0.01$. We show one stochastic solution for the stochastic model for the (left) urban and (right) rural patch; see Example \ref{ex:1}.} 
\end{figure}
\end{example}

\begin{example} \label{ex2}
We now consider
\begin{equation*}
	\begin{array}{ccccc}
\mm = 1/(365\cdot 80)\,, & \pu = 0.08\,, & \gu = 0.01\,, & \bu = 3\cdot 10^{-2}\,, & \Lambda_u = \mm N_{u0},\\
\mr = 1/(365\cdot 70)\,, & \pr = 0.40\,, & \gr = 0.10\,, & \br = 2\cdot 10^{-5}\,, & \Lambda_r = \mr N_{r0},\\.
	\end{array}
\end{equation*}
and movement in both directions $\dur=\dru=0.05$. For this set of parameters, $\mathcal{R}_{0u}\approx 2.9$, $\mathcal{R}_{0r}\approx 2\cdot 10^{-4}$, and the deterministic system admits the disease-free state and one stable endemic state given by $$(I_u^*/N_u^*, I_r^*/N_r^*) \approx (0.82, 0.76).$$ We remark that 
without population movement ($\dur=\dru=0$), the disease persists in the urban patch but dies out in the rural patch. For $\delta_{ur} = \delta_{ru} = 0.05$, we calculate that $\displaystyle \frac{\delta_{ur}}{\gamma_u+ \mu} \approx 5.98$, and conditions \eqref{eq:condR0} are satisfied. Therefore, the disease-free steady state is locally asymptotically stable; see Figure \ref{fig:example2_1234} where we use initial conditions 
\begin{equation*}
\begin{array}{ccc}
S_{u0} = 9999\,, & I_{u0} = 1\,, & \widetilde{S}_{u0} = 0\,,\\
S_{r0} = 3000\,, & I_{r0} = 0\,, & \widetilde{S}_{r0} = 0\,.
\end{array}
\end{equation*}

\begin{figure}[htb!]
\centering
\includegraphics[width=0.49\linewidth]{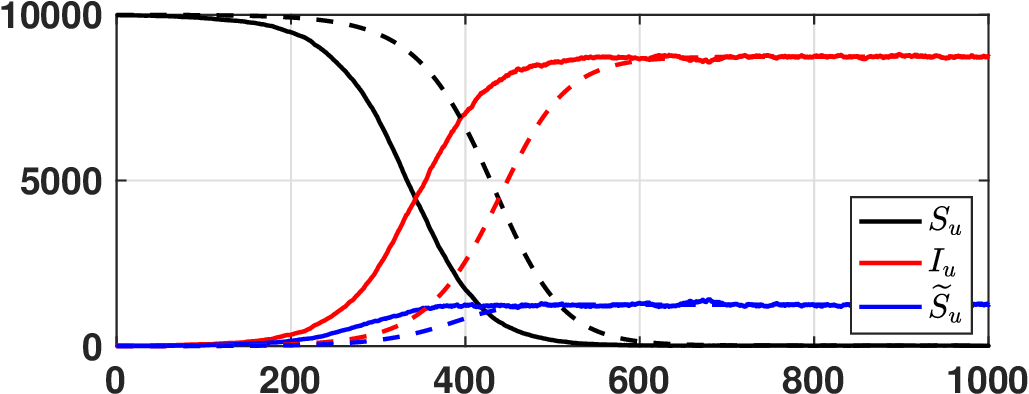}
\hfill
\includegraphics[width=0.49\linewidth]{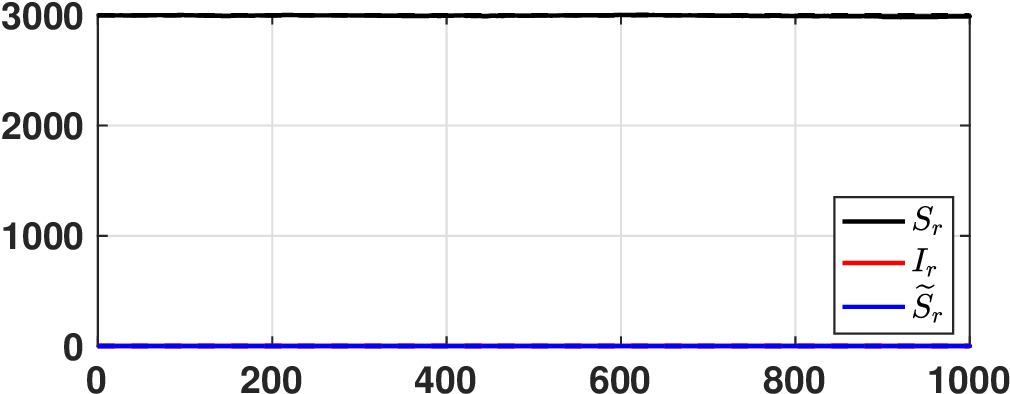}

\includegraphics[width=0.49\linewidth]{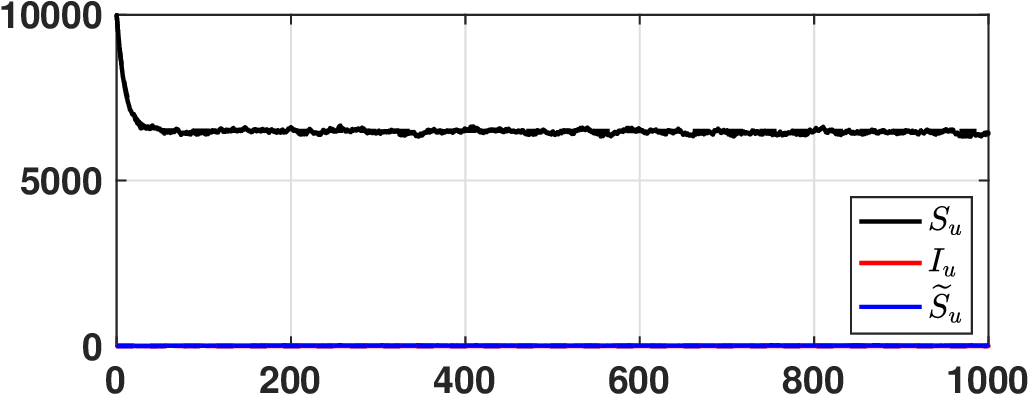}
\hfill
\includegraphics[width=0.49\linewidth]{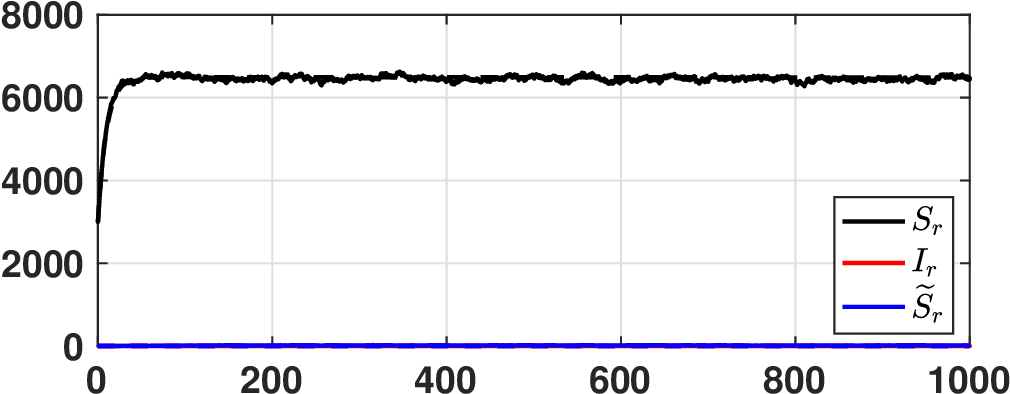}





\caption{\label{fig:example2_1234} Number of individuals for each compartment as a function of time (days) for (left) urban and (right) rural patches for the stochastic model. (Top) In the absence of population movement ($\delta_{ur} = \delta_{ru} = 0$), the urban patch reaches an endemic state. (Bottom) For $\delta_{ur} = \delta_{ru} = 0.05$, both populations reach a disease-free steady state. Population movement allows the disease to die out; see Example \ref{ex2}.}
\end{figure}

We then study the dependence on initial conditions:

\begin{enumerate}
\item When we increase slightly the number of infected individuals in the urban patch (while keeping constant the total population) to 
    \begin{equation*}
	\begin{array}{ccc}
    S_{u0} = 9990, & I_{u0} = 10\,, & \widetilde{S}_{u0} = 0\,,\\
    S_{r0} = 3000, & I_{r0} = 0\,, & \widetilde{S}_{r0} = 0\,,
	\end{array}
    \end{equation*}
    the disease-free steady state for the deterministic model \eqref{eq:sys} is locally asymptotically stable when there is movement between patches. This behavior is preserved for the stochastic models {and the three approximations} %
    as shown in see Figure~\ref{fig:ex2}. 
    \item When the number of initially infected individuals is sufficiently large, we reach the endemic steady state $(I_u^*/N_u^*, I_r^*/N_r^*)$; see Figure~\ref{fig:ex2c} for the case when $S_{u0}$ and $I_{u0}$ are changed to $S_{u0}=9000$, $I_{u0}=1000$ (keeping constant the total initial population), and Figure~\ref{fig:example2_56} for an illustration for the stochastic model with initial conditions $N_{u0} = 1000, I_{u0} \in \{1,100\}, \widetilde{S}_{u0} = 0, N_{r0} = 3000, I_{r0} = 0, \widetilde{S}_{r0} = 0$.
\end{enumerate}

Secondly, we slightly increase $\beta_u$ to $\beta_u = 5.3 \cdot 10^{-2}$, with a smaller population with similar percentages of initially infected individuals: 
\begin{equation*}
	\begin{array}{ccc}
S_{u0} = 999\,, & I_{u0} = 1\,, & \widetilde{S}_{u0} = 0\,,\\
S_{r0} = 300\,, & I_{r0} = 0\,, & \widetilde{S}_{r0} = 0\,.
\end{array}
\end{equation*}
Condition \eqref{lem:symsta_c} is no longer satisfied, and the system reaches an endemic steady state when we start even with only one infected individual in the urban patch; see Figure \ref{fig:ex2d}. 


\begin{figure}[ht!]
\centering
\includegraphics[width=0.49\linewidth]{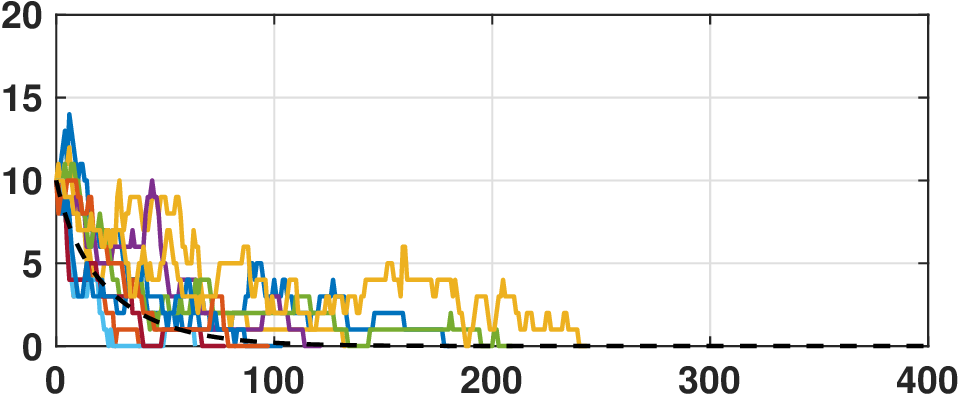} 
\includegraphics[width=0.49\linewidth]{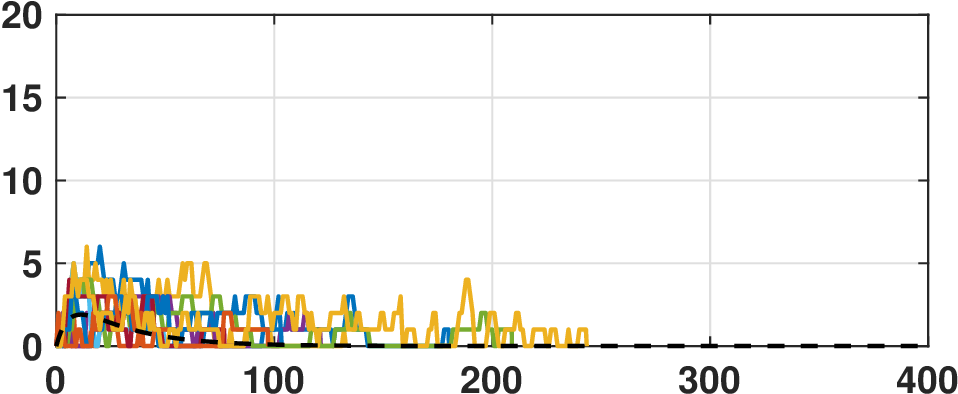} 
\includegraphics[width=0.49\linewidth]{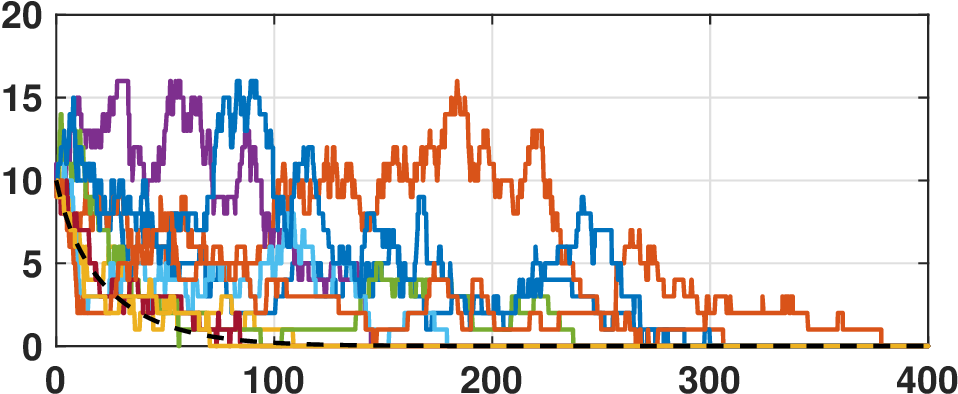} 
\includegraphics[width=0.49\linewidth]{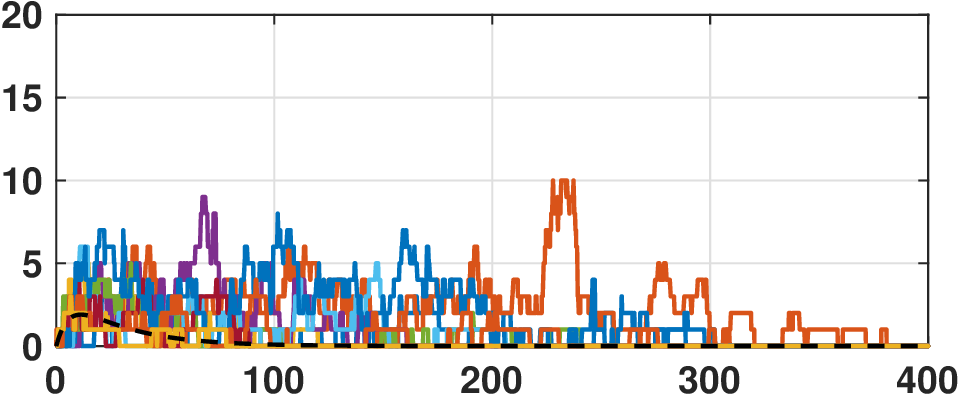} 
\includegraphics[width=0.49\linewidth]{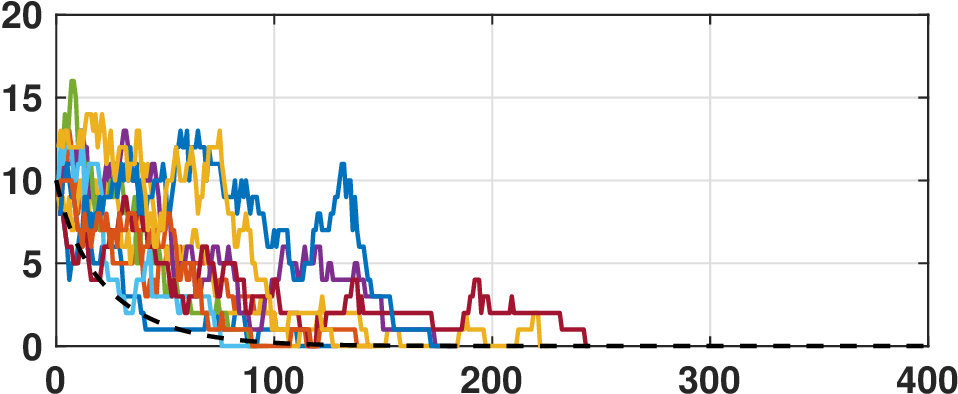} 
\includegraphics[width=0.49\linewidth]{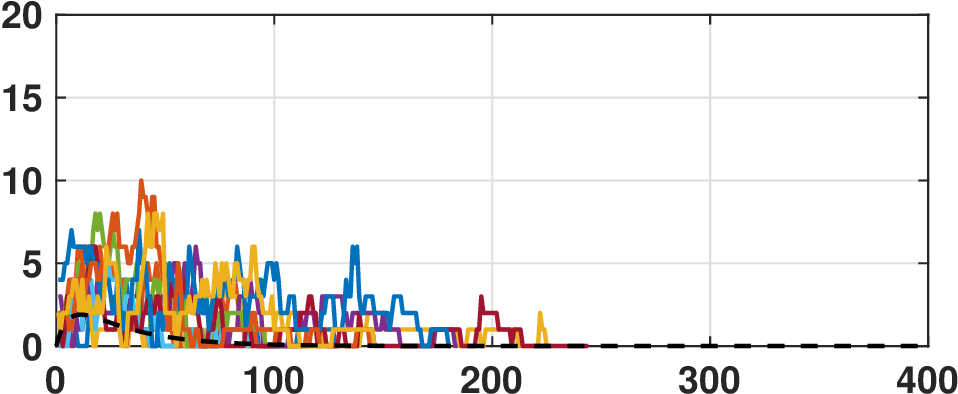} 
\includegraphics[width=0.49\linewidth]{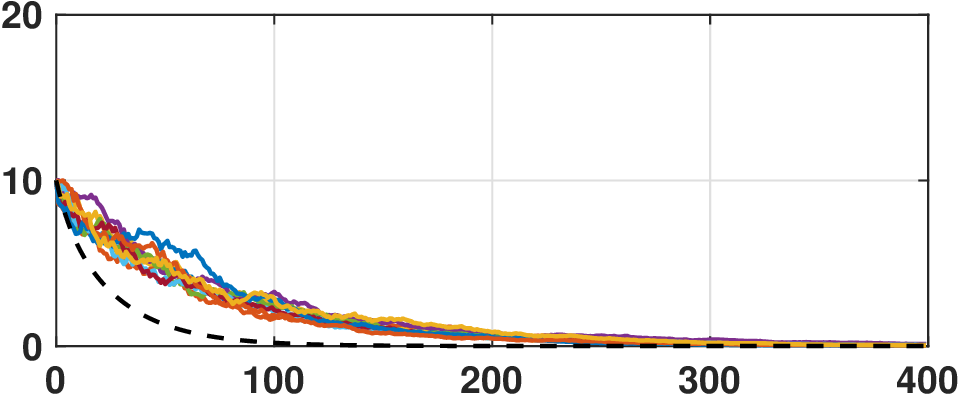}
\includegraphics[width=0.49\linewidth]{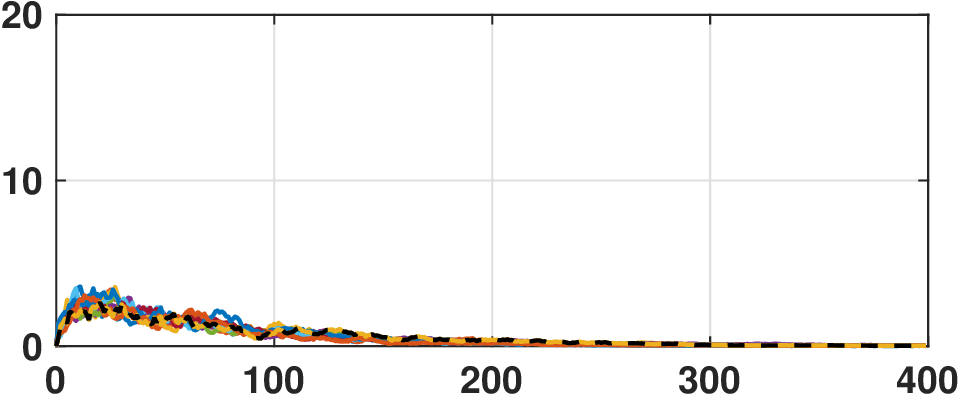}
\caption{\label{fig:ex2}(left) $I_u$ and (right) $I_r$ as a function of time (in days) for ten simulations with $\dur = 0.05$, $\dru = 0.05$ for the stochastic, DTMC, Poisson and SDE models. The deterministic solution is shown in the black dashed line; see Example \ref{ex2}.}
\end{figure}

\begin{figure}[htb!]
\centering
\includegraphics[width=0.49\linewidth]{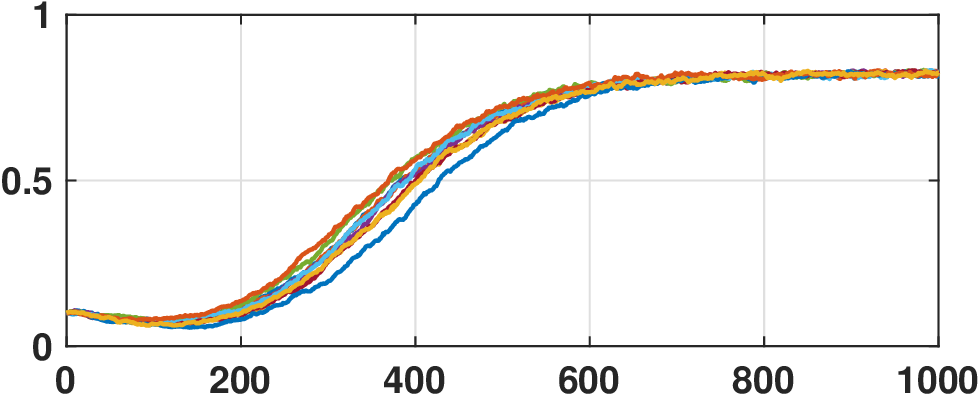} 
\hfill
\includegraphics[width=0.49\linewidth]{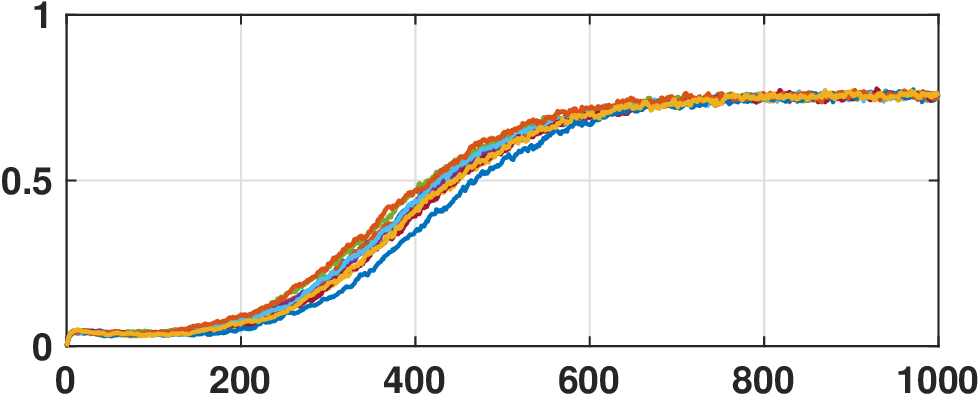} 
\caption{\label{fig:ex2c}(left) $I_u/N_u$ and (right) $I_r/N_u$ as a function of time (in days) for ten simulations with $\dur = 0.05$, $\dru = 0.05$ for the stochastic model, with $I_{u,0}=1000$. The deterministic solution is shown in the black dashed line; see Example \ref{ex2}. The curves for the three approximations of the stochastic model behave similarly and are omitted.}
\end{figure}

\begin{figure}[htb!]
\centering
\includegraphics[width=0.49\linewidth]{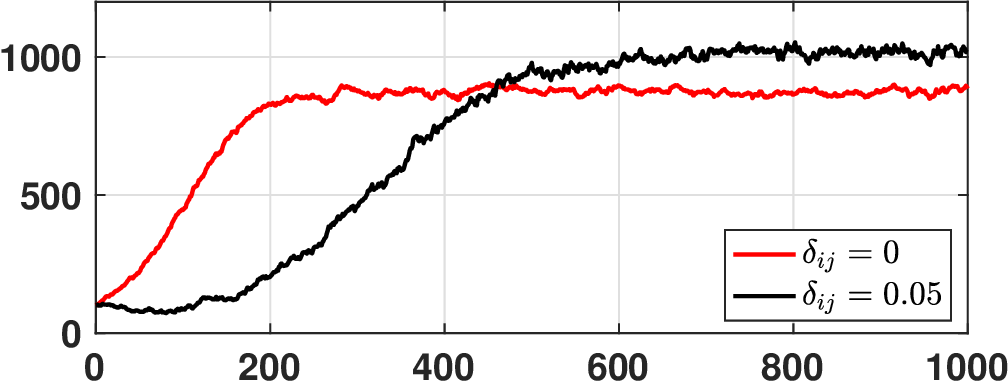} 
\hfill
\includegraphics[width=0.49\linewidth]{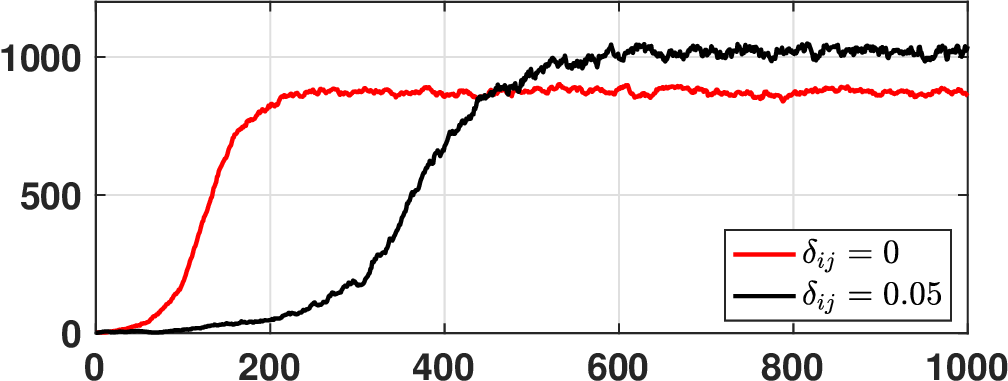} 
\caption{\label{fig:example2_56} Total infected population ($I_u+I_r$) as a function of time (days) for the stochastic model, with $\delta_{ur}=\delta_{ru}=0$ and $\delta_{ur}=\delta_{ru}=0.05$. We show the effect of the conditions in \eqref{eq:condR0} on the class of infected individuals; see Example \ref{ex2}. We use (left) $I_{u0}=100$ and (right) $I_{u0}=1$. {The curves for the three approximations of the stochastic model behave similarly
and are omitted.} }  
\end{figure}

\begin{figure}[htb!]
\centering
\includegraphics[width=0.49\linewidth]{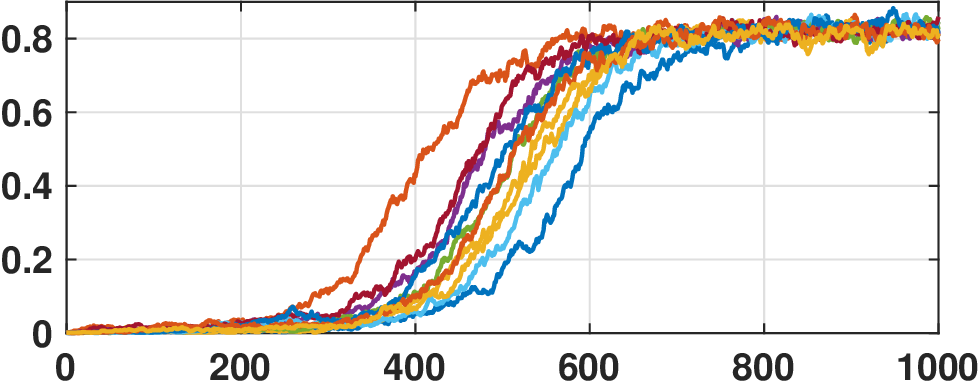} 
\hfill
\includegraphics[width=0.49\linewidth]{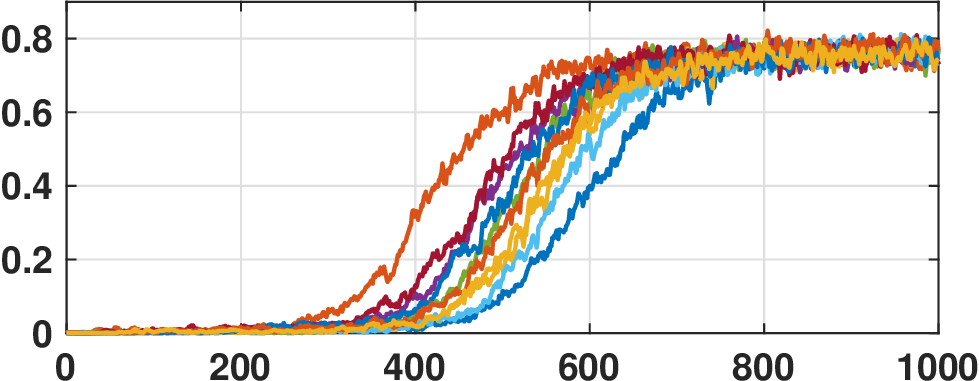} 
\caption{\label{fig:ex2d}(left) $I_u/N_u$ and (right) $I_r/N_u$ as a function of time (in days) for ten simulations with $\dur = 0.05$, $\dru = 0.05$ for the stochastic model, with $\beta_u = 5.3\cdot 10^{-2}$, $R_{u0}\approx 5.28$, $I_{u0}=1$. The deterministic solution is shown in the black dashed line; see Example \ref{ex2}. Of a total of 1000 simulations, 805 reach a free disease state.
}
\end{figure}

\end{example}

\subsection{Stochastic source-sink dynamics}

Stochasticity promotes disease extinction in different ways. Here, we considered two cases. First, one of the populations has an $ R_0$ value close to one. Therefore, small outbreaks may take place but followed by disease extinction with a high probability. Second, we consider that one patch is below the critical community size, and then, even for $R_0>1$, disease extinction will eventually occur. In both cases, one patch has $R_0>1$, and its population is well above the critical community size; therefore, an endemic equilibrium is reached. Those patches act as a source of infection. In the two examples analyzed below, the simulations of the stochastic (Table \ref{table:transitions}) and the SDE models produce qualitatively different results.

\begin{example} \label{ex:9} 

First, consider a situation where $R_{0u} > 1$ and $R_{0r} \approx 1$ taking into account the following initial conditions and parameters:

\begin{equation*}
\begin{array}{ccccc}
S_{u0} = 99990\,, & I_{u0} = 10\,, & \widetilde{S}_{u0} = 0\,, \\
S_{r0} = 29990\,, & I_{r0} = 10\,, & \widetilde{S}_{r0} = 0\,, \\
\mu_u = 1/(50 \cdot 365)\,, & \gamma_u  =1/10\,, & R_{0u} = 1.5\,,  \\
\mu_r=1/(50 \cdot 365)\,, &  \gamma_r =1/10\,, & R_{0u} = 1.05\,,\\
\rho_u = 0.7 \cdot \beta_u\,, & \Lambda_u = \mu_u N_{u0} \,, & \delta_{ur} = 0.000001 \\
\rho_r = 0.7 \cdot \beta_r\,, & \Lambda_r = \mu_r N_{r0} \,, & \delta_{ru} = 0.000001 N_{u0}/N_{r0}, \\
\end{array}
\end{equation*}
\noindent considering in both cases that $\beta_x = R_{0x} (\gamma_x + \mu_x)$ with $x \in \{u,r\}$. 

We have source-sink dynamics for this set of parameters, whereas, in the urban patch, we have an endemic situation that acts as the source for the rural patch. There are no differences between the stochastic model and its approximations in the urban patch, as seen in Figure \ref{fig:example9}. However, in the rural patch, the stochastic model predicts micro-outbreaks followed by the extinction of the infected population. On the other hand, the stochastic differential equations model produces an endemic regime without the extinction of the disease for long periods. 

\begin{figure}[htb!]
\centering
\includegraphics[width=\linewidth]{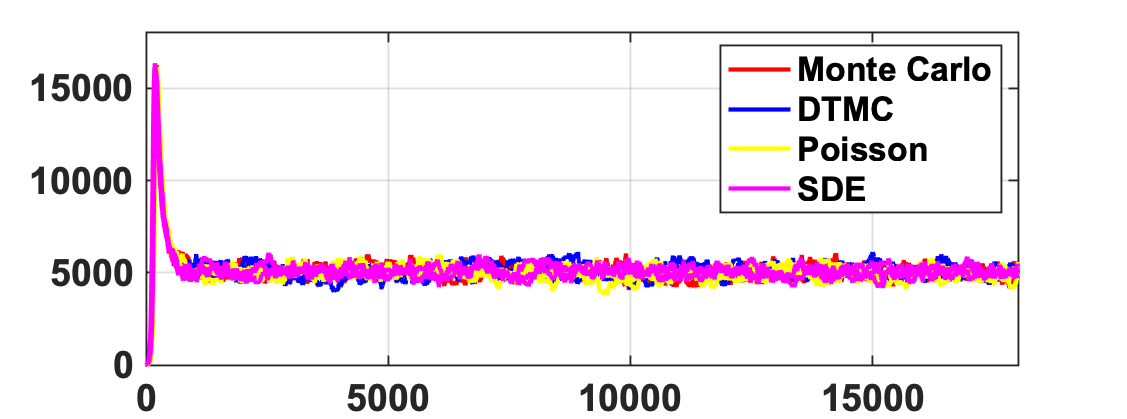}
\hfill
\includegraphics[width=0.49\linewidth]{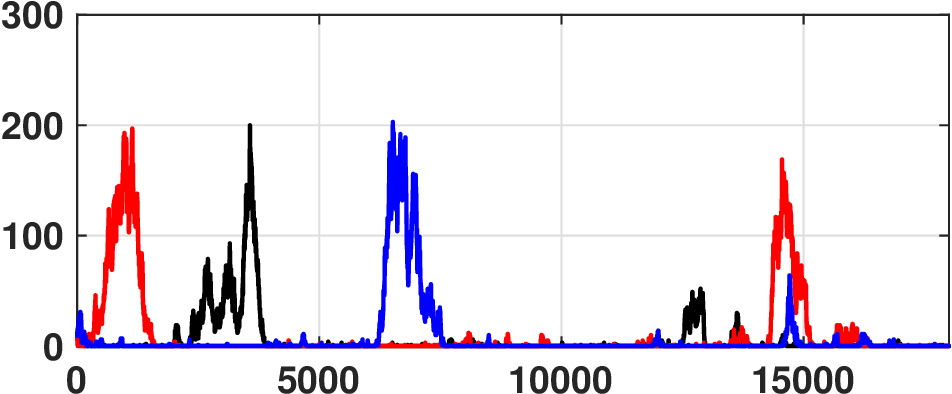}
\includegraphics[width=0.49\linewidth]{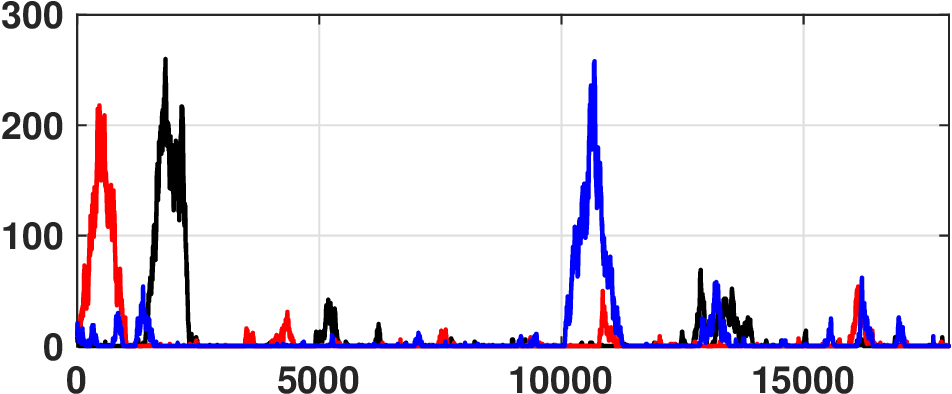}

\includegraphics[width=0.49\linewidth]{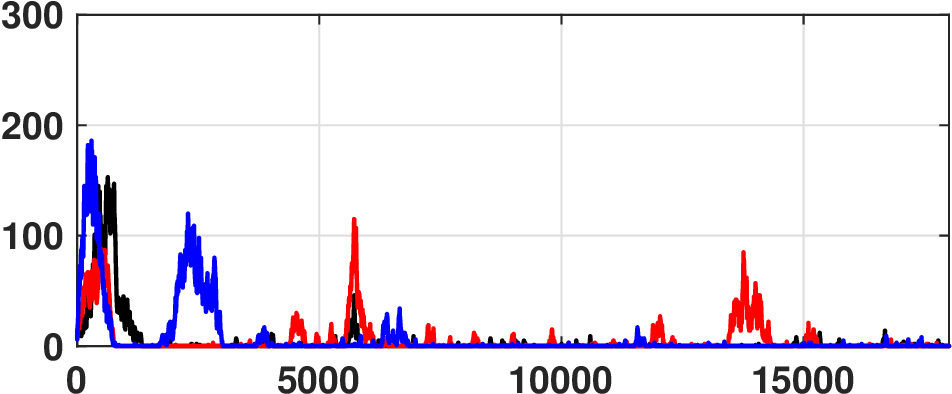}
\includegraphics[width=0.49\linewidth]{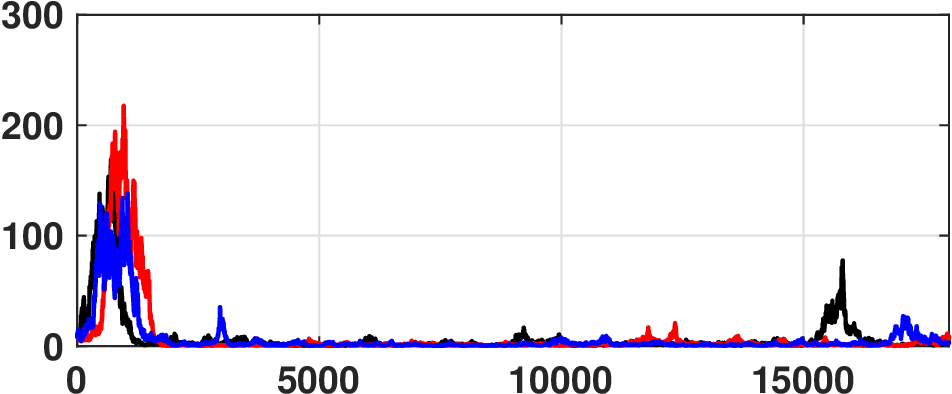}

\caption{\label{fig:example9} Number of infected individuals in the (top) urban patch for all methods, and three simulations of the rural populations as a function of time (in days) considering the stochastic model (middle left), DTMC scheme (middle right), Poisson scheme (bottom left) and SDE scheme (bottom right); see Example \ref{ex:9}.}
\end{figure}


We considered the number of days without infected individuals to quantify the differences between the dynamics in the rural patch. Considering the continuous SDE approach, we count the days with less than one infected individual. We carried out 1000 simulations of 70 simulated years for each numerical scheme. We obtained that, on average, the number of days without infected individuals is equal to 19648 days considering the stochastic model simulations, 19852 in the case of the Discrete Time Markov Chain, and 19701 for the Poisson scheme. In the case of SDE, we do not have days without infected individuals, but we have an average of 8646 days with less than one infected individual, a significant difference. The boxplots of these results are displayed in Figure \ref{fig:example9_boxplot}. 



\begin{figure}[htb!]
\centering
\includegraphics[width=\linewidth]{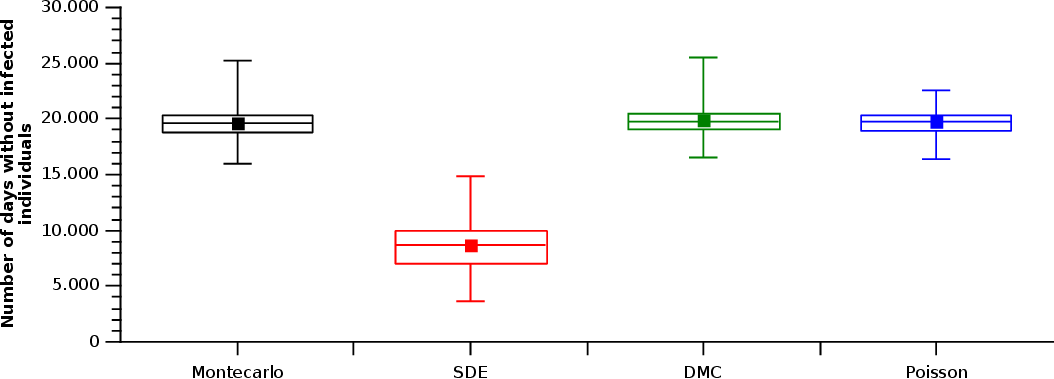}
\caption{\label{fig:example9_boxplot} Number of days without infected individuals considering the different schemes. For the SDE scheme, we considered the number of days with less than one infected individual; see Example \ref{ex:9}.}
\end{figure}

\end{example}

\FloatBarrier

\begin{example} \label{ex:10}

Another situation that produces source-sink dynamics is generated when the population in one patch is close (or below) to the critical community size (CCS) while the other is well above the CCS and acts as a source of infection. CCS is the minimum size of a closed population that may sustain an endemic equilibrium. Below the CCS, the disease dies out after a major outbreak \cite{bartlett1957,bartlett1960}. To simulate this situation, we consider the following parameters, 

\begin{equation*}
\begin{array}{ccccc}
N_{u0} = 10^5\,, & I_{u0} = 10\,, & \widetilde{S}_{u0} = 0\,, \\
N_{r0} = 10^4\,, & I_{r0} = 10\,, & \widetilde{S}_{r0} = 0\,, \\
\mu_u = 1/(50 \cdot 365)\,, & \gamma_u  =1/5\,, & R_{0u} = 1.5\,,  \\
\mu_r=1/(50 \cdot 365)\,, &  \gamma_r =1/5\,, & R_{0u} = 1.5\,,\\
\rho_u = 0.675 \cdot \beta_u\,, & \Lambda_u = \mu_u N_{u0} \,, & \delta_{ur} = 0.000001 \\
\rho_r = 0.675 \cdot \beta_r\,, & \Lambda_r = \mu_r N_{r0} \,, & \delta_{ru} = 0.000001 N_{u0}/N_{r0}, \\
\end{array}
\end{equation*}

\noindent where $\beta_x = R_{0x} (\gamma_x + \mu_x)$ with $x \in \{u,r\}$. 

The populations in each patch have the same parameter values, and the only difference is the size of the populations. Figure \ref{fig:example10} shows a typical realization obtained with the different stochastic models. 

\begin{figure}[htb!]
\centering
\includegraphics[width=\linewidth]{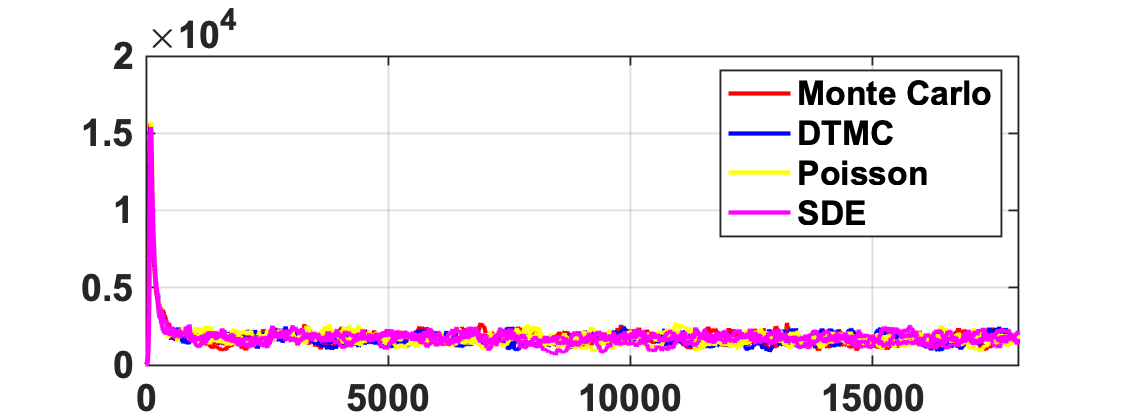}
\hfill
\includegraphics[width=0.49\linewidth]{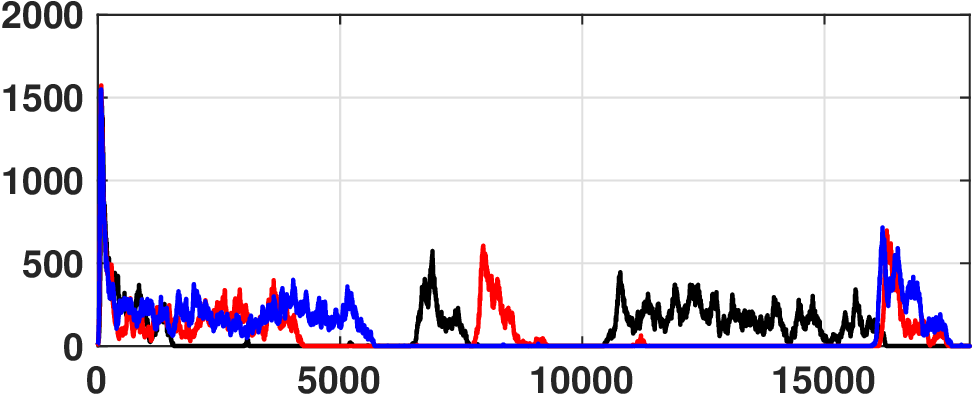}
\includegraphics[width=0.49\linewidth]{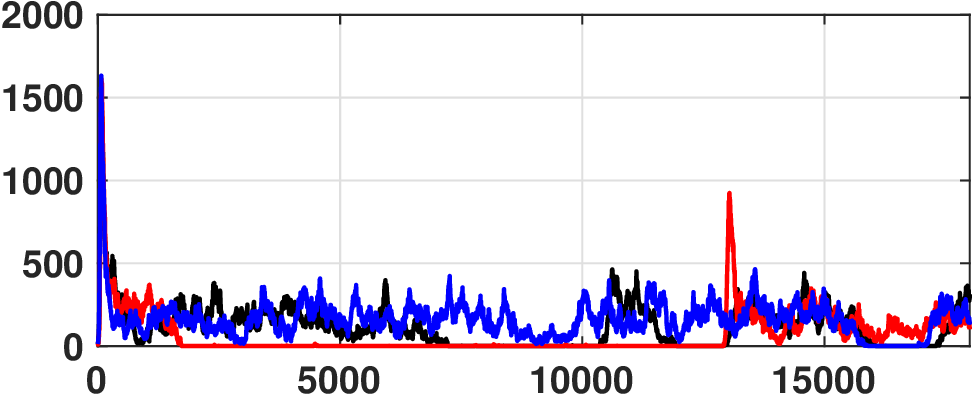}
\includegraphics[width=0.49\linewidth]{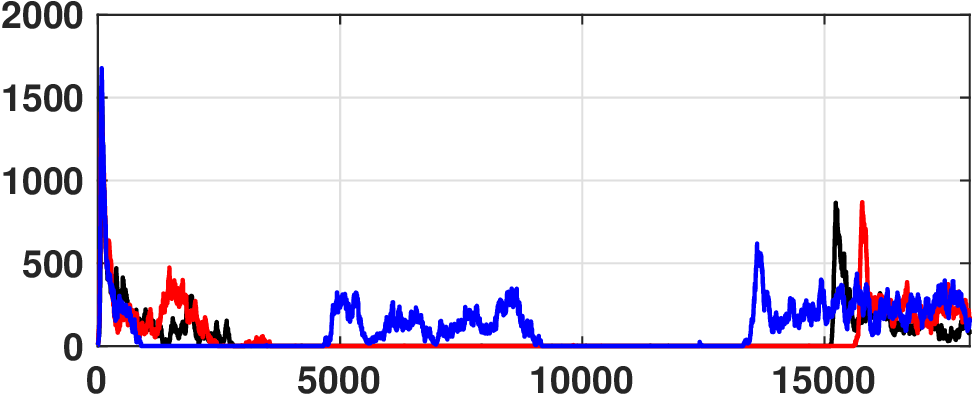}
\includegraphics[width=0.49\linewidth]{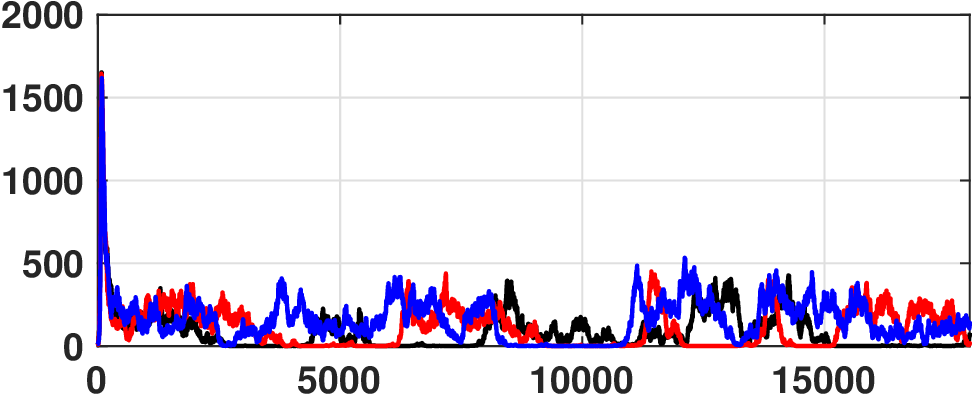}

\caption{\label{fig:example10} 
Number of infected individuals in the (top) urban patch for all methods, and three simulations of the rural populations as a function of time (in days) considering the stochastic model (middle left), DTMC scheme (middle right), Poisson scheme (bottom left) and SDE scheme (bottom right); see Example \ref{ex:10}.}
\end{figure}

As before, we can quantify the differences in the dynamics by counting the number of days without infected individuals in each numerical scheme. In this sense, we can see that the differences are significant (Figure \ref{fig:example10_boxplot}), showing qualitatively different dynamics in the rural patch when the SDE scheme is considered. 

\begin{figure}[htb!]
\centering
\includegraphics[width=\linewidth]{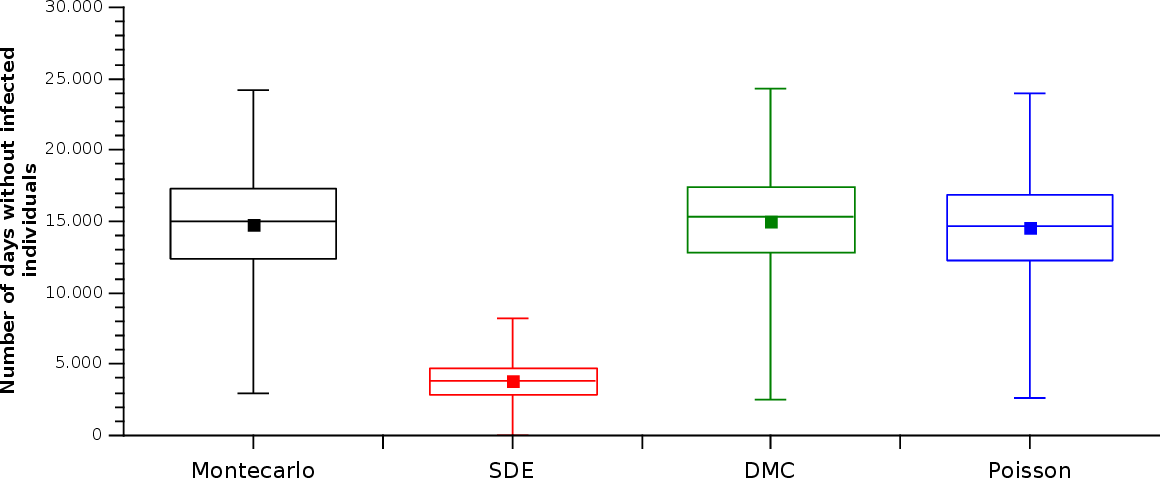}
\caption{\label{fig:example10_boxplot} Number of days without infected individuals considering the different schemes. For the SDE scheme, we considered the number of days with less than one infected individual; see Example \ref{ex:10}.}
\end{figure}

\end{example}


\section{Conclusion} \label{sec:conclusions}

This study has developed a framework for understanding the dynamics of infectious diseases across heterogeneous patches using a stochastic model incorporating nonlinear recidivism. By applying three distinct stochastic approximations (discrete-time Markov chain, Poisson, and stochastic differential equations), the research illustrates how stochasticity can critically influence the trajectory and outcome of epidemics in ways that deterministic models might not fully capture.

One of the major findings of this work is the significant role that population movement plays in shaping epidemic dynamics. The migration between urban and rural patches affects the spread rate and the persistence of the disease within each patch. Notably, our models show that even low levels of mobility between patches can prevent disease extinction in a patch where the disease would otherwise die out, illustrating the critical impact of human movement on disease control and elimination strategies.

Furthermore, our results indicate that stochastic effects can lead to qualitative differences in disease dynamics, particularly in scenarios near critical thresholds such as the community size or basic reproduction number, $\mathcal{R}_0$. In such cases, stochastic models predict extinction events and oscillations in disease prevalence that deterministic models fail to capture. These findings emphasize the necessity of considering stochastic models in public health planning, especially for diseases with high extinction probability or those affected by frequent small outbreaks.

In the particular case of stochastic differential equations, we had to deal with the problem of positivity, also considering that the extinction of the disease in a patch does not imply the end of the transmission of the disease. On the other hand, we seek that the numbers drawn have a biological meaning, seeking that they are always non-negative. In simpler systems of stochastic differential equations, this problem can be solved using logarithmic transformations (see \cite{chen2021} and cites therein); however, applying this approach in a system with so many equations was too complex and beyond the objectives of the work. For this reason, we decided to use truncated normal distributions for its numerical implementation, inspired by what was proposed by \cite{krause2018}. In this sense, we noticed that the simulations in which truncated normal distributions controlled the positivity were more similar to the deterministic simulations compared to those with no control over the positivity of the quantities drawn.

Moreover, while our models provide valuable insights, their computational complexity and the need for extensive data to parameterize them might limit their practical use in real-time epidemic surveillance and response. Future research should focus on simplifying the models without significantly compromising accuracy, developing hybrid models that can dynamically switch between deterministic and stochastic frameworks based on the phase of the epidemic, and incorporating more realistic social dynamics and behavioral changes.

This study contributes a deeper understanding of how stochastic processes influence epidemic dynamics across spatially structured populations. It highlights the importance of stochasticity in epidemic modeling and offers a robust framework for exploring intervention strategies in a heterogeneous landscape. Furthermore, enhancing the models' realism and applicability to real-world scenarios will be crucial for leveraging these insights in public health decision-making.





\section*{acknowledgments} J.G.C. and F.S. thank the Research Center in Pure and Applied Mathematics and the School of Mathematics at Universidad de Costa Rica for their support during the preparation of this manuscript.

\appendix

\section{Covariance for the SDE system} \label{ap:expectedValueCovariance}

Let $\Delta \bx(t) = (\Delta S_u, \Delta I_u,\Delta \widetilde{S}_u, \Delta S_r, \Delta I_r,\Delta \widetilde{S}_r)^T$. 
The covariance matrix $$\mathbb{V}\big(\Delta \bx(t)\big)=\mathbb{E}\big(\Delta \bx(t) \;\Delta \bx(t)^{T}\big)-\mathbb{E}\big(\Delta \bx(t)\big) \mathbb{E}\big(\Delta \bx(t)\big)^{T}$$ is approximated by dropping the second order terms, for which elements are of order $(\Delta t)^2$. Hence, 

\noindent\makebox[\textwidth]{\begin{minipage}{\dimexpr\eqboxwidth{Eq}+1cm\relax}
\medskip
\begin{align*}
\mathbb{V}&\big(\Delta \bx(t)\big) \approx \mathbb{E}\big(\Delta \bx(t) \;\Delta \bx(t)^{T}\big)\\
&=\mathbb{E} \begin{bmatrix}
(\Delta S_u)^2 & \Delta S_u\, \Delta I_u & \Delta S_u\, \Delta \widetilde{S}_u & \Delta S_u\, \Delta S_r & \Delta S_u\, \Delta I_r & \Delta S_u\, \Delta \widetilde{S}_r \\
\Delta I_u\, \Delta S_u & (\Delta I_u)^2 & \Delta I_u\, \Delta \widetilde{S}_u & \Delta I_u\, \Delta S_r & \Delta I_u\, \Delta I_r & \Delta I_u\, \Delta \widetilde{S}_r \\
\Delta \widetilde{S}_u\, \Delta S_u & \Delta \widetilde{S}_u\, \Delta I_u & (\Delta \widetilde{S}_u)^2 & \Delta \widetilde{S}_u\, \Delta S_r & \Delta \widetilde{S}_u\, \Delta I_r & \Delta \widetilde{S}_u\, \Delta \widetilde{S}_r \\
\Delta S_r\, \Delta S_u & \Delta S_r\, \Delta I_u & \Delta S_r\, \Delta \widetilde{S}_u & (\Delta S_r)^2 & \Delta S_r\, \Delta I_r & \Delta S_r\, \Delta \widetilde{S}_r \\
\Delta I_r\, \Delta S_u & \Delta I_r\, \Delta I_u & \Delta I_r\, \Delta \widetilde{S}_u & \Delta I_r\, \Delta S_r & (\Delta I_r)^2 & \Delta I_r\, \Delta \widetilde{S}_r \\
\Delta \widetilde{S}_r\, \Delta S_u & \Delta \widetilde{S}_r\, \Delta I_u & \Delta \widetilde{S}_r\, \Delta\widetilde{S}_u & \Delta \widetilde{S}_r\, \Delta S_r & \Delta \widetilde{S}_r\, \Delta I_r & (\Delta \widetilde{S}_r)^2
\end{bmatrix}.
\end{align*}
\end{minipage}}\vspace{\belowdisplayskip}
After straightforward computations, we obtain that the diagonal entries of the covariance matrix are given by
\begin{align*}
& V_{11} = \delta_{ur}\, S_u + \delta_{ru}\, S_r + \frac{\beta_u\, I_u\, S_u}{N_u} + \Lambda_u + \mu_u S_u, \\
& V_{22} = \delta_{ur}\, I_u + \delta_{ru}\, I_r + \frac{\beta_u\, I_u\, S_u}{N_u} + \frac{\rho_u\, I_u\, \widetilde{S}_u}{N_u} + \gamma_u\, I_u + \mu_u\, I_u, \\
& V_{33} = \delta _{ur}\, \widetilde{S}_u + \delta _{ru}\, \widetilde{S}_r + \frac{\rho_u\, I_u\, \widetilde{S}_u}{N_u} + \gamma_u\, I_u + \mu_u\, \widetilde{S}_u, \\
& V_{44} = \delta_{ur}\, S_u + \delta_{ru}\, S_r + \frac{\beta_r\, I_r\, S_r}{N_r} + \Lambda_r + \mu_r\, S_r\\
& V_{55} = \delta_{ur}\, I_u + \delta_{ru}\, I_r + \frac{\beta_r\, I_r\, S_r}{N_r} + \frac{\rho_r\, I_r\, \widetilde{S}_r}{N_r} + \gamma_r\, I_r + \mu_r\, I_r, \\
& V_{66} = \delta _{ur}\, \widetilde{S}_u + \delta _{ru}\,  \widetilde{S}_r + \frac{\rho_r\, I_r\, \widetilde{S}_r}{N_r} + \gamma_r\, I_r  + \mu_r\, \widetilde{S}_r,
\end{align*}
and the remaining non-zero entries of the covariance matrix are
\begin{align*}
    V_{12}&=V_{21}=-\frac{\beta_u\, I_u\, S_u}{N_u},\\
    V_{41}&=V_{14}=-\dur S_u-\dru S_r, \\
    V_{23}&=V_{32}=-\frac{\rho_u\, I_u\, \widetilde{S}_u}{N_u} - \gamma_u\, I_u,\\ V_{25}&=V_{52}=-\dur I_u-\dru I_r, \\
    V_{36}&=V_{63}=-\delta _{ur}\, \widetilde{S}_u - \delta _{ru}\, \widetilde{S}_r, \\
    V_{45}&=V_{54}=-\frac{\beta_r\, I_r\, S_r}{N_r} , \\
    V_{56}&=V_{65}=-\frac{\rho_r\, I_r\, \widetilde{S}_r}{N_r} - \gamma_r I_r
\end{align*}

We can rewrite the above formulas using matrix notation. Define

$$C = 
\begin{bmatrix}
-1 & 0 & 0 & 1 & 0 & 0 \\
 1 & 0 & 0 &-1 & 0 & 0 \\ 
 0 &-1 & 0 & 0 & 1 & 0 \\
 0 & 1 & 0 & 0 &-1 & 0 \\
 0 & 0 &-1 & 0 & 0 & 1 \\ 
 0 & 0 & 1 & 0 & 0 &-1 \\ 
-1 & 1 & 0 & 0 & 0 & 0 \\
 0 & 1 &-1 & 0 & 0 & 0 \\ 
 0 &-1 & 1 & 0 & 0 & 0 \\ 
 1 & 0 & 0 & 0 & 0 & 0 \\ 
-1 & 0 & 0 & 0 & 0 & 0 \\ 
 0 &-1 & 0 & 0 & 0 & 0 \\
 0 & 0 &-1 & 0 & 0 & 0 \\ 
 0 & 0 & 0 &-1 & 1 & 0 \\
 0 & 0 & 0 & 0 & 1 &-1 \\
 0 & 0 & 0 & 0 &-1 & 1 \\ 
 0 & 0 & 0 & 1 & 0 & 0 \\ 
 0 & 0 & 0 &-1 & 0 & 0 \\ 
 0 & 0 & 0 & 0 &-1 & 0 \\
 0 & 0 & 0 & 0 & 0 &-1
\end{bmatrix}
\quad \text{and} \quad
p = 
\begin{bmatrix}
\delta _{ur}\, S_u \\
\delta _{ru}\, S_r\\
\delta _{ur}\, I_u\\
\delta _{ru}\, I_r\\
\delta _{ur}\, \widetilde{S}_u\\
\delta _{ru}\, \widetilde{S}_r\\
\frac{\beta_u\, I_u\, S_u}{N_u}\\ 
\frac{\rho_u\, I_u\, \widetilde{S}_u}{N_u}\\
\gamma_u\, I_u\\\
\Lambda_u\\
\mu_u\, S_u\\
\mu_u\, I_u \\ 
\mu_u\, \widetilde{S}_u\\
\frac{\beta_r\, I_r\, S_r}{N_r}\\ 
\frac{\rho_r\, I_r\, \widetilde{S}_r}{N_r}\\
\gamma_r\, I_r\\
\Lambda_r\\
\mu_r\, S_r\\
\mu_r\, I_r\\\
\mu_r\, \widetilde{S}_r
\end{bmatrix},$$
where $C$ is the matrix with the changes for all possible events, and $p$ is the vector with corresponding transition rates as in Table \ref{table:transitions}. Then, we can write
$$\mathbb{E}\big(\Delta \bx(t)\big) = C^T p\ \Delta t,$$
and
$$\mathbb{V}\big(\Delta \bx(t)\big) \approx \mathbb{E}\big(\Delta \bx(t) \;\Delta \bx(t)^{T}\big) = V \Delta t,$$
where $D = \text{diag}(p)$ is a diagonal matrix whose diagonal is $p$ and $V = C^T D C$. We can write $V=G G^T$ where the non-zero entries of the matrix $G$ are given by



\[
\begin{array}{llll}
G_{1,1} = -\sqrt{\dur S_u}, & G_{4,1} = \sqrt{\dur S_u}, & G_{1,2} = \sqrt{\dru S_r},\\
G_{4,2} = -\sqrt{\dru S_r}, & G_{2,3} = -\sqrt{\dur I_u}, & G_{5,3} = \sqrt{\dur I_u},\\
G_{2,4} = \sqrt{\dru I_r}, & G_{5,4} = -\sqrt{\dru I_r}, &
G_{3,5} = -\sqrt{\dur \widetilde{S}_u},\\
G_{6,5} = \sqrt{\dur \widetilde{S}_u}, & G_{3,6} = \sqrt{\dru \widetilde{S}_r}, &  G_{6,6} = -\sqrt{\dru \widetilde{S}_r}, \\
G_{1,7} = -\sqrt{\beta_u \dfrac{I_u}{N_u} S_u}, & G_{2,7} = \sqrt{\beta_u \dfrac{I_u}{N_u} S_u}, & G_{2,8} = \sqrt{\rho_u \dfrac{I_u}{N_u} \widetilde{S}_u},\\
G_{3,8} = -\sqrt{\rho_u \dfrac{I_u}{N_u} \widetilde{S}_u}, &
G_{2,9} = -\sqrt{\gamma_u I_u}, & G_{3,9} = \sqrt{\gamma_u I_u},\\
G_{1,10} = \sqrt{\Lambda_\mu}, & G_{1,11} = -\sqrt{\mu_u S_u}, & G_{2,12} = -\sqrt{\mu_u I_u},\\
G_{3,13} = -\sqrt{\mu_u \widetilde{S}_u}, & G_{4,14} = -\sqrt{\beta_r \dfrac{I_r}{N_r} S_r}, & G_{5,14} = \sqrt{\beta_r \dfrac{I_r}{N_r} S_r},\\
G_{5,15} = \sqrt{\rho_r \dfrac{I_r}{N_r} \widetilde{S}_r}, & G_{6,15} = -\sqrt{\rho_r \dfrac{I_r}{N_r} \widetilde{S}_r}, & G_{5,16} = -\sqrt{\gamma_r I_r},\\
G_{6,16} = \sqrt{\gamma_r I_r}, & G_{4,17} = \sqrt{\Lambda_{\mr}}, & G_{4,18} = -\sqrt{\mr S_r},\\
G_{5,19} = -\sqrt{\mr I_r}, & G_{6,20} = -\sqrt{\mr \widetilde{S}_r}.
\end{array}
\]


\section{Numerical scheme for solving the SDE} \label{ap:numerical_scheme}

As seen on the right side of the SDE system (Sect. \ref{sec:sde}), each event comprises the average value plus a random value. For example, in the case of deaths we have $death_{X_j}(t) = \mu_j X_j(t) \Delta t + \sqrt{\mu_j X_j(t) \Delta t} dW$, where $i,j \in \{u,r\}$ with $i \neq j$, $X \in \{S,I, \widetilde{S}\}$. In this formulation, if the mean value ($\mu_j X_j(t) \Delta t$) is close to 0, then the resulting value ($death_{X_j}(t)$) has a non-negligible probability of being negative. The same situation occurs in the rest of the events involved in the SDE system. This means that the drawn values may not have a correct biological meaning.

To avoid this situation, we consider an approach similar to the one proposed by~\cite{krause2018}, using truncated normal random variables. Given a normal random variable $T \sim N(\mu, \sigma)$, the truncated normal distribution of $T$ between $t_1$ and $t_2$ has the following density function,

\begin{equation*}
g(t; \mu, \sigma, t_1, t_2) = \left\lbrace
\begin{array}{ll}
\displaystyle \frac{1}{\sigma} \frac{f(\frac{t - \mu}{\sigma})}{F(\frac{t_2 - \mu}{\sigma}) - F(\frac{t_1 - \mu}{\sigma})} & \textup{if }  t_1 \leq t \leq t_2 \\
0 & \textup{otherwise}
\end{array}
\right.
\end{equation*}

\noindent where $f$ and $F$ are the density and distribution functions of a standard normal random variable.  

Therefore, using that density function and considering $\mu = \mu_j X_j(t) \Delta t$, $\sigma = \sqrt{\mu_j X_j(t) \Delta t}$, $t_1 = 0$ and $t_2 = 2\mu_j X_j(t) \Delta t$ we can generate random values for deaths, which are centered on $\mu_j X_j(t) \Delta t$, have a symmetric distribution, and are non-negative. It is important to note that as the mean moves away from 0, the truncated normal distribution approaches the normal distribution. A similar procedure can be used for other events.

Let $R(\mu, \sigma)$ be the random number generated from a normal random variable with mean $\mu$ and deviation $\sigma$ truncated between $0$ and $2\mu$. The following numerical scheme was used to solve the SDE presented in Section \ref{sec:sde}. At each time step $\Delta t$, we calculate

\begingroup
\allowdisplaybreaks
\begin{align}
birth_j(t) & = R \left( \Lambda_j \Delta t,\sqrt{\Lambda_j \Delta t}\right), \label{eq:nacu}\\
death_{X_j}(t) & = R\left(\mu_j X_j(t) \Delta t, \sqrt{\mu_j X_j(t) \Delta t}\right), \\
new\_inf_j(t) & = R \left( \frac{\beta_j S_j(t) I_j(t)}{N_j(t)} \Delta t , \sqrt{\frac{\beta_j S_j(t) I_j(t)}{N_j(t)} \Delta t} \right),
 \\
new\_reinf_j(t) & = R\left( \frac{\rho_j \widetilde{S}_j(t) I_j(t)}{N_j(t)} \Delta t,  \sqrt{ \frac{\rho_j \widetilde{S}(t) I_j(t)}{N_j(t)} \Delta t} \right), \\
new\_rec_j & =R\left(\gamma_j I_j \Delta t, \sqrt{\gamma_j I_j \Delta t}\right), \\
migX_{ij} & = R\left(\delta_{ij} S_i(t) \Delta t, \sqrt{\delta_{ij} S_i(t) \Delta t} \right) \label{eq:migRru},
\end{align}
\endgroup
where $i,j \in \{u,r\}$ with $i \neq j$, $X \in \{S,I, \widetilde{S}\}$, and $\Lambda_j = \mu_j Neq_j$. We then compute the values at time $t+\Delta t$ with the following formulas:

\begingroup
\allowdisplaybreaks
\begin{align*}
S_u(t+\Delta t) = S_u(t)+&birth_u(t)-death_{S_u}(t)-new\_inf_u(t) \\
 & - migS_{ur}(t) + migS_{ru}(t), \nonumber\\
I_u(t+\Delta t)  = I_u(t)+&new\_inf_u(t)-new\_rec_u(t)+new\_reinf_u(t)-death_{I_u}(t)\\
&-migI_{ur}(t)+migI_{ru}(t), \nonumber\\
\widetilde{S}_u(t+\Delta t)=R_u(t)+&rec_u(t)-new\_reinf_u(t)-death_{\widetilde{S}_u}(t)\\
&-migR_{ur}(t)+migR_{ru}(t), \nonumber\\
S_r(t+\Delta t) = S_r(t)+&birth_r(t)-death_{S_r}(t)-new\_inf_r(t) \\
&- migS_{ru}(t) + migS_{ur}(t), \nonumber\\
I_r(t+\Delta t) =I_r(t)+&new\_inf_r(t)-new\_rec_r(t)+new\_reinf_r(t)-death_{I_r}(t)\\
&-migI_{ru}(t)+migI_{ur}(t), \nonumber\\
\widetilde{S}_r(t+\Delta t)=\widetilde{S}_r(t)+&new\_rec_r(t)-new\_reinf_r(t)-death_{\widetilde{S}_r}(t)\\
&-migR_{ru}(t)+migR_{ur}(t),
\end{align*}
\endgroup
and the total population for each patch is updated as
\begingroup
\allowdisplaybreaks
\begin{align*}
N_u(t+\Delta t)& =S_u(t+\Delta t)+I_u(t+\Delta t)+\widetilde{S}_u(t+\Delta t), \nonumber\\
N_r(t+\Delta t)& =S_r(t+\Delta t)+I_r(t+\Delta t)+\widetilde{S}_r(t+\Delta t). \nonumber
\end{align*}
\endgroup

At each time step $\Delta t$, we control that equations (\ref{eq:nacu} - \ref{eq:migRru}) are not greater than the allowed values. In the case in which that condition is not fulfilled, we set this variable in the maximum value, i.e., 

\begin{itemize}
\item If $death_{X_j}(t)>X_j(t) $, then $ death_{X_j}(t)=X_j(t) $    
\item If $new\_inf_j>S_j $, then $ new\_inf_j=S_j$  
\item If $new\_reinf_j> \widetilde{S}_j $, then $ new_reinf_j=\widetilde{S}_j$  
\item If $new\_rec_j>I_j $, then $ new\_rec_j=I_j;$
\item If $ migX_{ij} > X_i $, then $ migX_{ij}=X_i;$

\end{itemize}

In addition, at each time step $\Delta t$, if $X(t + \Delta t) < 0$, we set $X(t + \Delta t) = 0$.

\end{document}